\documentclass{article}
\usepackage[utf8]{inputenc}
\usepackage{setspace}

\usepackage[margin=1in]{geometry}

\usepackage{apacite}
\usepackage{blindtext}
\usepackage{amsmath,bm}
\usepackage{color}
\usepackage{rotating}

\usepackage{caption}
\usepackage{subcaption}

\usepackage{grffile}

\usepackage{graphicx}

\usepackage{lscape}

\usepackage{blindtext}

\usepackage{amssymb}
\usepackage{varioref} 
\usepackage{textcomp}

\usepackage{booktabs}

\usepackage{multirow}

\usepackage{pdflscape}

\usepackage{graphicx}

\usepackage{relsize}
\usepackage{tablefootnote}
\usepackage{wrapfig}
\usepackage{natbib}
\usepackage{listings}
\usepackage{longtable}
\usepackage{url}
\usepackage{tabularx}
\usepackage{float}
\usepackage{epsfig}
\usepackage[titletoc]{appendix}
\usepackage{arydshln}
\usepackage{footnote}
\usepackage{secdot}
\usepackage{bigints}
\usepackage[utf8]{inputenc}
\usepackage[english]{babel}
\usepackage{xr}
\usepackage{adjustbox}
\usepackage{graphicx}

\usepackage{multicol}
\usepackage{enumitem}

\usepackage{bigstrut}
\usepackage{rotating}
\usepackage{xr}

\title{Destructive cure models with proportional hazards lifetimes and associated likelihood inference}
\author{N. Balakrishnan\thanks{N. Balakrishnan is with Department of Mathematics and Statistics, McMaster University, Hamilton, Ontario, Canada, L8S 4K1 (email: bala@mcmaster.ca).} \hspace{0.7mm} and S. Barui \thanks{Corresponding Author: S. Barui is with Quantitative Methods and Operations Management Area, Indian Institute of Management, Kozhikode, India (email: sandipbarui@iimk.ac.in).}}
\date{\today}

\begin{document}
\maketitle
\begin{abstract}
In survival analysis, cure models have gained much importance due to rapid advancements in medical sciences. More recently, a subset of cure models, called destructive cure models, have been studied extensively under competing risks scenario wherein initial competing risks undergo a destructive process, such as under a chemotherapy. In this article, we study destructive cure models by assuming a flexible weighted Poisson distribution (exponentially weighted Poisson, length biased Poisson and negative binomial distributions) for the initial number of competing causes and with lifetimes of the susceptible individuals following proportional hazards.  The expectation-maximization (EM) algorithm and profile likelihood approach are made use of for estimating the model parameters.  An extensive simulation study is carried out under various parameter settings to examine the properties of the models, and the accuracy and  robustness of the proposed estimation technique. Effects of model mis-specification on the parameter estimates are also discussed in detail. Finally, for  the illustration of the proposed methodology, a real-life cutaneous melanoma data set is analyzed.               
\end{abstract}


\section{Introduction}\label{sect1}

\hspace{5mm} Classical modeling techniques in survival analysis, such as proportional hazards model or accelerated failure time model, assume every individual under study will inevitably experience the event of interest (e.g., death, relapse, etc.). However, owing to outstanding progress in medical science over the past few decades, there often exists a fraction of subjects who do not encounter the event of interest or recurrences, even when followed-up long enough. These individuals are termed as cured or immune or long-term survivors in the literature, and models in lifetime data analysis consisting of a cure fraction are called cure models. Thus, the population under study consists of a mixture of subjects, {\it viz.}, cured and susceptible (non-cured). If we assume $I$ to be a binary random variable (r.v.) indicating cured if $I=0$ and susceptible if $I=1$, then the population survival function $S_p(y)$ can be expressed as 
\begin{flalign}\label{eqx1}
S_p(y) = P(Y>y) &=\sum_{i=0}^1 P(Y>y|I=i)P(I=i) =\pi_0+(1-\pi_0)S_u(y),
\end{flalign}      
where $Y$ denotes time to event, $\pi_0=P(I=0)$ represents the cured fraction or rate,  and $S_u(y)=P(Y>y|I=1)$ denotes the survival function of susceptible individuals. Note that $S_u(y)$ is a proper survival function whereas $S_p(y)$ is not, since $\lim_{y \to \infty}S_p(y)=\pi_0$.  The application of cure models extends beyond survival analysis to many applied fields (\citealp*{maller1996survival}). One of the primary objectives is to estimate $\pi_0$ along with lifetime parameters of such cure models. However, estimating $\pi_0$ can be complicated, and the estimates are difficult to validate as the individuals who are cured are also censored, and the information on all censored individuals is missing. Further, the definition of `cure' is implicit, varies across studies, and not always well defined. \\

\hspace{5mm} One of the earliest developments can be found in the work of \cite{boag1949maximum} in which he introduced the cure model emphasizing on the information loss in conventional five year survival rate from a clinician's viewpoint. \cite{berkson1952survival} estimated the cure fraction using a least-squares method by considering a mixture cure model, followed by \cite{haybittle1965two} who estimated the proportion of treated cancer patients surviving to a specific time with respect to the normal population. \cite{farewell1982use} mapped covariates using a logit-link to the proportion of susceptible and the lifetime distribution as Weibull. \cite{larson1985mixture} considered the failure types to be distributed as multinomial and the conditional distribution of the failure time, given the type, as piecewise exponential under proportional hazards. A semi-parametric generalization of lifetime distribution of the susceptible individuals, as suggested by \cite{farewell1982use}, was proposed by \cite{kuk1992mixture} by introducing a Cox's proportional hazards model, where the regression parameters were estimated by maximizing a Monte Carlo approximation of marginal likelihood and baseline survival function by the EM algorithm (\citealp{dempster1977maximum}). \cite{sy2000estimation} further worked with the proportional hazards model, but used a Breslow-type and product-limit estimators for estimating the baseline hazard and baseline survival functions, respectively;  \cite{peng2000nonparametric} improved on the estimation technique under similar assumptions of the model. A consideration of bounded cumulative hazard function was made by \cite{tsodikov2003estimating} as an alternative to a two-component mixture model and provided an extension to the proportional hazards regression model. As opposed to the mixture cure models, \cite{chen1999new} proposed a promotion time cure model considering the underlying biological process generating the failure times based on a Poisson distribution (\citealp{yakovlev1993stochastic, yakovlev1996stochastic}). By considering a Box-Cox transformation on the population survival function, an intermediate model between a promotion time and mixture cure model was discussed by \cite{yin2005cure}. A class of semi-parametric transformation models, with proportional hazards and proportional odds promotion time cure models, as special cases was suggested by \cite{zeng2006semiparametric}. A theoretical study on the existence, consistency and asymptotic normality of the semi-parametric maximum likelihood (ML) estimator under proportional hazards set-up was done by \cite{fang2005maximum}. \cite{rodrigues2009poisson} introduced a flexible COM-Poisson cure model under competing risks scenario, and this model was studied subsequently by \cite{balakrishnan2012algorithm, balakrishnan2013expectation, balakrishnan2014algorithm}.\\

\hspace{5mm} More pragmatic alternative to cure models, called destructive cure models, was introduced by  \cite{rodrigues2011destructive} with the assumption that the initial competing causes undergo a process of destruction.  Specifically, let  $M$  be the initial number of latent competing causes related to the event of interest. In cancer studies, often the event of interest is patient's death which can be caused by one or more malignant metastasis-component (\citealp{yakovlev1996stochastic}) tumor cells. After chemotherapy or radiation, only a portion of initial metastasis-component cells remain active and undamaged, thereby, reducing the initial number of competing causes. Given $M=m$, we may consider $X_g$ as a Bernoulli random variable (r.v.), distributed independently of $M$, which takes 1 if the $g$-th competing cause is present (i.e., if the $g$-th malignant tumor cell remains undamaged after the treatment) with probability $p \in (0, 1)$ and 0 otherwise.  Thus, if we define
\begin{equation} 
D = \begin{cases} X_1+ \ldots+X_M &, \mbox{if } M>0, \\ 
                                 0 &, \mbox{if } M=0, \end{cases}
\end{equation}
then $D$ represents the number of initial competing causes that are not destroyed. Obviously, $D \le M$ and the conditional distribution of $D$, given $M=m$, is known as the damaged distribution that follows a binomial distribution with parameters $m$ and $p$ if $m>0$, and $P(D=0|M=0)=1$. \cite{rodrigues2011destructive} discussed the destructive cure model by considering the distribution of $M$ as weighted Poisson. The probability mass function (pmf) of $M$ following a weighted Poisson distribution is given by
\begin{equation} 
P(M=m; \eta, \phi)=  \begin{cases} \frac{\Omega(m; \phi)}{\mathbb{E}_{\eta}[\Omega(M; \phi)]} p^*(m; \eta) &, m=0, 1, 2, \dots,  \\  0     &,\mbox{ otherwise,} \end{cases}
\end{equation}
where $\Omega(.; \phi)$ is a non-negative weight function characterized by $\phi$ with $\phi \in \mathbb{R}$, $p^*(.; \eta)$ is the pmf of a Poisson distribution with parameter $\eta > 0$, and  $\mathbb{E}_{\eta}[.]$ is expectation taken with respect to the Poisson pmf.   \cite{gallardo2016algorithm} developed an EM algorithm based technique for the same model for estimating the parameters for the three special cases, {\it viz.}, destructive length-biased Poisson, destructive exponentially weighted Poisson and destructive negative binomial cure  models. An extension of this model was described by \cite{borges2012correlated} by creating a correlation structure between the initiated cells using the generalized power series distribution. A Bayesian method of inference was further proposed in the context of destructive weighted Poisson cure model by \cite{rodrigues2012bayesian}. Interested readers may refer to  \cite{cancho2013destructive} and \cite{pal2015likelihood,  pal2016destructive, pal2016likelihood} for some further discussions in this regard.\\

\hspace{5mm} Here, we consider the initial number of competing causes $M$ to have a weighted Poisson distribution with weight function $\Omega(m; \phi)$ as $m$, $e^{\phi m}$ and $\Gamma (m+\phi^{-1})$, following \cite{pal2015likelihood,  pal2016destructive, pal2016likelihood}. The corresponding models are known as destructive length-biased Poisson (DLBP), destructive exponentially weighted Poisson (DEWP) and destructive negative binomial (DNB) cure models, respectively. Given $D=d$, let $W_j$ (latent) be the time-to-event associated with the $j$-th competing cause. Now, $W_j$s are assumed to be independently and identically distributed (i.i.d.) with common cumulative distribution function (cdf) $F(w)$ and common survival function $S(w)=1-F(w)=P(W_j > w)$, for all $j=1, \ldots, d$. In order to accommodate the proportion of individuals who will not encounter the event of interest, we introduce a degenerate r.v. $W_0$ such that $P(W_0=\infty)=1$. Thus, we only observe $Y=\min\{W_0, W_1, \dots, W_D\}$ and the population survival function in this case is given by $$S_p(y)=P(Y>y)=\sum_{d=0}^\infty P(D=d) \{S(y)\}^d=G_D(S(y)),$$
where $G_D(.)$ is the probability generating function of $D$ evaluated at $S(y)$. The population density function is defined as $f_p(y)=-\frac{dS_p(y)}{dy}$. Here, $f_p(.)$ is not a proper density function, but plays a significant role in likelihood-based estimation of model parameters. The novelty of the present work lies in the fact that the model considers a proportional hazards structure with 
\begin{equation}\lambda(w;\bm \gamma, \bm x)=\lambda_0(w)e^{\bm \gamma' \bm x},\end{equation} 
where $\lambda(w; \bm \gamma, \bm x)$ and $\lambda_0(w)$ are the respective hazard and baseline hazard functions related to $W_j$, for all $j=1, \dots, d$, $\bm x$ is a covariate vector with corresponding parameter vector $\bm \gamma$ of same dimension. By assuming proportional hazards structure, lifetime of each subject $i$ is linked to the specific characteristic $\bm x_i, i=1, \dots, n$, intrinsic to that  individual. Thus, this model is a generalization of the one considered by \cite{pal2016destructive}, which assumes lifetime distribution to be the same across all subjects in the study. Further, the model considered here enables us to verify non-homogeneity in lifetime distribution across individuals by conducting a hypothesis test for $H_0: \bm \gamma=0$ vs. $H_a:\bm \gamma \ne 0$. Recently, \cite{balakrishnan2019destructive} have studied destructive cure model with non-homogeneous lifetimes by assuming a proportional odds structure under the  cure rate scenario. Owing to the flexibility and robustness imparted by the shape and scale parameters, a Weibull distribution is widely used in survival analysis to model lifetimes,  and for this reason we consider the baseline hazard function $\lambda_0(w)$ to be a Weibull hazard function. \\

\hspace{5mm} The rest of the paper proceeds as follows. Model assumptions and formulation are explained in Section \ref{sect2}. The form of the data and the likelihood function are described in Section \ref{sect3}, while the method of estimation of model parameters using EM algorithm and computation of asymptotic standard error (SE) of the estimates of parameters are provided in Section \ref{sect4}. The proposed model and inferential methods are applied to a real-life data set obtained from a cutaneous melanoma study, and this is discussed in Section \ref{sect5}. An extensive simulation study is carried out with various settings for examining the accuracy of the estimation method, and the pertinent details are given in Section \ref{sect6}. A model discrimination is performed and discussed in Section \ref{sect7}. Finally, some concluding remarks are made in Section \ref{sect8}.


\section{Model description}\label{sect2}
\subsection{Cure models}\label{sect2_1}
\hspace{5mm} As mentioned in the last section, $M$ is assumed to follow a weighted Poisson distribution with three candidate weight functions $m$, $e^{\phi m}$ and $\Gamma(m + \phi^{-1})$. The corresponding three models possess the following forms and properties. 
\subsubsection{Destructive length-biased Poisson cure model}\label{subsect2.1.1}
\hspace{5mm}Assuming $\Omega(m; \phi)=m$, the pmf of $M$ is given by 
\begin{equation} 
P(M=m; \eta, \phi)=  \begin{cases} \frac{e^{-\eta} \eta^{m-1}}{(m-1)!} &, m= 1, 2, \dots  \\  0     &,\mbox{ otherwise,} \end{cases}
\end{equation}
which is a shifted Poisson distribution, shifted by 1. Because $(D|M=m) \sim $ Bernoulli($m$, $p$), the unconditional pmf of $D$, i.e., the number of active competing causes, is given by 
\begin{flalign}
P(D=d; \eta, \phi, p) = & \sum_{m=d}^\infty P(D=d|M=m)P(M=m)\nonumber   \\
                                 =& \sum_{m=d}^\infty \frac{m!}{(m-d)!d!}p^d(1-p)^{m-d}\frac{e^{-\eta} \eta^{m-1}}{(m-1)!}\nonumber \\
                                 =& \frac{e^{-\eta p}(\eta p)^d}{d!}\left(1-p+ \frac{d}{\eta}\right),\text{  } d=0, 1, 2, ... 
\end{flalign}
The cure rate is then given by \begin{equation}\label{eqxx7.5}\pi_0 = P(D=0)=e^{-\eta p}(1-p),\end{equation} while the population survival  and  density functions are   
\begin{equation}\label{eqx8}
S_p(y)= e^{-\eta p F(y)}\{1-p F(y)\}
\end{equation}
and
\begin{equation}\label{eqx9}
f_p(y) = \eta p f(y)e^{-\eta p F(y)}\left\{1-p F(y)-\frac{pf(y)}{\eta}\right\},
\end{equation}
where $f(.)$ is the common probability density function (pdf) of $W_j$, for all $j=1, 2, \dots, d$.

\subsubsection{Destructive exponentially weighted Poisson cure model}\label{subsect2.1.2}
\hspace{5mm}Under this model, we assume $\Omega(m; \phi)=e^{\phi m}$ as the weight function, which gives the pmf of $M$ as
\begin{equation} 
P(M=m; \eta, \phi)=  \begin{cases}e^{-\eta e^{\phi}}\frac{(\eta e^{\phi})^m}{m!} &, m= 0, 1, 2, \dots  \\  0     &,\mbox{ otherwise.} \end{cases}
\end{equation}
This is readily seen to be a Poisson distribution with rate parameter $\eta e^{\phi}$. The unconditional distribution of the undamaged number of initial competing causes $D$ is given by
\begin{flalign}\label{eqx10}
P(D=d; \eta, \phi, p)=& \sum_{m=d}^\infty \frac{m!}{(m-d)!d!}p^d(1-p)^{m-d}e^{-\eta e^{\phi}}\frac{(\eta e^{\phi})^m}{m!}\nonumber \\
                                 = &  e^{-\eta p e^{\phi}}\frac{(\eta p e^{\phi})^d}{d!},\text{  } d=0, 1, 2, ..., 
\end{flalign}
which is a Poisson distribution with rate parameter $\eta p e^{\phi}$. Consequently,
\begin{equation}\label{eqxx9.5}
\pi_0=e^{-\eta p e^{\phi}},
\end{equation}
\begin{equation}\label{eqx10.1}
S_p(y)=e^{-\eta p e^{\phi}F(y)}
\end{equation}
and
\begin{equation}\label{eqx11}
f_p(y)=\eta p e^{\phi}f(y)e^{-\eta p e^{\phi}F(y)}.
\end{equation}

Note that the model reduces to a destructive Poisson cure model if $\phi=0$. Furthermore, the choice of  $p=1$ in (\ref{eqx10}) yields Poisson or promotion-time cure model. 

\subsubsection{Destructive negative binomial cure model}\label{subsect2.1.3}
\hspace{5mm}Let us consider 
\begin{equation} \label{eqx14}
P(M=m; \eta, \phi)=  \begin{cases} \frac{\Gamma(m+\phi^{-1})}{\Gamma \phi^{-1}m!}\left(\frac{\phi \eta}{1+ \phi \eta}\right)^m (1+\phi \eta)^{-\phi^{-1}} &, m= 0, 1, 2, \dots  \\  0     &,\mbox{ otherwise,} \end{cases}
\end{equation}
where $M$ is a negative binomial r.v. with parameters $\phi^{-1}$ and  $\frac{\phi \eta}{1+ \phi \eta}$.  This is also a weighted Poisson distribution with parameter $\frac{\phi \eta}{1+ \phi \eta}$ and the weight function $\Omega(m; \phi)=\Gamma(m+ \phi^{-1})$, $\phi >0$.
Hence, the unconditional pmf of $D$ is given by 
\begin{flalign}\label{eqx15}
P(D=d; \eta, \phi, p) =& \sum_{m=d}^\infty P(D=d|M=m)P(M=m)\nonumber   \\
                                 = & \frac{p^d}{d!} \left(\frac{\phi \eta}{1+\phi \eta}\right)^d (1+\phi \eta)^{-\phi^{-1}} \sum_{m=d}^{\infty}\frac{\Gamma(m+ \phi^{-1})}{(m-d)!\Gamma(\phi^{-1})} \left[\frac{(1-p) \phi \eta}{1+\phi \eta}\right]^{m-d} \nonumber \\
&= \frac{\Gamma(d+\phi^{-1})}{\Gamma \phi^{-1}d!}\left(\frac{p \phi \eta}{1+ p \phi \eta}\right)^d (1+p\phi \eta)^{-\phi^{-1}},\text{  } d=0, 1, 2, ...
\end{flalign}
Evidently,  $D$ has a negative binomial distribution with parameters  $\phi^{-1}$ and $\frac{p\phi \eta}{1+ p\phi \eta}$. The corresponding cure rate, population survival function and population density function are given by  
\begin{equation}\label{eqx16}
\pi_0=(1+p \eta \phi)^{-\phi^{-1}},
\end{equation}
\begin{equation}\label{eqx17}
S_p(y)=(1+p \eta \phi F(y))^{-\phi^{-1}}
\end{equation}
and
\begin{equation}\label{eqx18}
f_p(y)=\eta pf(y) (1+p \eta \phi F(y))^{-(\phi^{-1}+1)}.
\end{equation}
Note that this destructive negative binomial cure model includes destructive geometric ($\phi=1$), negative binomial ($p=1$) and geometric ($\phi=1$ and $p=1$) cure models all as special cases.  \\

\subsection{Modeling lifetimes}\label{subsect2.2}
\hspace{5mm}Given $D=d$, we assume the hazard function $\lambda(.; \bm x, \bm z, \bm \gamma_2, \bm \gamma_3)$ of $W_j$ ($j=1, \dots, d$) to follow proportional hazards structure, i.e.,
\begin{equation}\label{eqxx20}
\lambda(w; \bm x, \bm z, \bm \gamma_2, \bm \gamma_3) = \lambda_0(w)e^{\bm \gamma_2' \bm x+\bm \gamma_3' \bm z},
\end{equation}
where  $\bm \gamma_2=(\gamma_{21}, \dots, \gamma_{2q_1})' \in \mathbb{R}^{q_1}$, $\bm \gamma_3=(\gamma_{31}, \dots, \gamma_{3q_2})'\in \mathbb{R}^{q_2}$, and $\bm x$ and $\bm z$ are respectively $q_1$ and $q_2$ dimensional covariate vectors, with $\lambda_0(w)$ being the baseline hazard function which does not depend on the covariates $\bm x$ and $\bm z$. The reasons for splitting the covariates into two parts, $\bm x$ and $\bm z$, and for considering $\bm \gamma_2$ and $\bm \gamma_3$ without the intercept terms are discussed in Section \ref{sect3}. In the literature, $\lambda_0(.)$ has been estimated by assuming either some well-known fully parametric distributions (\citealp{farewell1982use, balakrishnan2013lognormal, balakrishnan2013expectation, pal2016likelihood}), or by non-parametric methods (\citealp{kuk1992mixture, sy2000estimation, fotios}). In this article, we assume $\lambda_0(w)$ to be a Weibull hazard function of the form
$$\lambda_0(w)=\lambda_0(w; \gamma_0, \gamma_1)=\gamma_0 {\gamma_1}^{-\gamma_0} {w}^{\gamma_0-1},  w>0, $$ where $\gamma_0>0$ and $\gamma_1>0$ are the respective shape and scale parameters of the Weibull distribution. The Weibull distribution is closed under proportional hazards when the shape parameter remains constant. Moreover, a two-parameter Weibull distribution provides a great degree of flexibility to the lifetimes of susceptible individuals since it includes cases of decreasing ($\gamma_0 < 1$), constant ($\gamma_0=1$, i.e., exponential distribution) and increasing ($\gamma_0 > 1$) failure rates. Let us denote $\bm \gamma=(\gamma_0, \gamma_1, \bm \gamma'_2, \bm \gamma'_3)'$. Thence, 
\begin{equation}\label{eqx21}
\lambda(w; \bm x, \bm z, \bm \gamma)= \gamma_0 \left(\gamma_1 e^{-\frac{\bm \gamma_2' \bm x+ \bm \gamma_3' \bm z}{\gamma_0}}\right)^{-\gamma_0} w^{\gamma_0-1}
\end{equation}
is also a Weibull hazard function with shape $\gamma_0$ and scale $\gamma_1e^{-\frac{\bm \gamma_2' \bm x+\bm \gamma_3' \bm z}{\gamma_0}}$. Consequently, for $w>0$, 
\begin{flalign}\label{eqx22}
S(w)=S(w; \bm x, \bm z, \bm \gamma)&=\exp \left\{-w^{\gamma_0}\left(\gamma_1e^{-\frac{\bm \gamma_2' \bm x+\bm \gamma_3' \bm z}{\gamma_0}}\right)^{-\gamma_0} \right\},
\end{flalign}
\begin{flalign}\label{eqx23}
F(w)=F(w; \bm x, \bm z, \bm \gamma)&=1-\exp \left\{-w^{\gamma_0}\left(\gamma_1e^{-\frac{\bm \gamma_2' \bm x+\bm \gamma_3' \bm z}{\gamma_0}}\right)^{-\gamma_0} \right\}
\end{flalign}
and
\begin{flalign}\label{eqx24}
f(w)=f(w; \bm x, \bm z, \bm \gamma)=\gamma_0 \left(\gamma_1 e^{-\frac{\bm \gamma_2' \bm x+ \bm \gamma_3' \bm z}{\gamma_0}}\right)^{-\gamma_0} w^{\gamma_0-1} \exp \left\{-w^{\gamma_0}\left(\gamma_1e^{-\frac{\bm \gamma_2' \bm x+\bm \gamma_3' \bm z}{\gamma_0}}\right)^{-\gamma_0} \right\}
\end{flalign}
are the corresponding common cdf, survival function and pdf of each $W_j$, $j=1, \dots, d$.\\

\hspace{5mm} \cite{pal2016destructive, pal2016likelihood} proceeded by assuming the lifetime distribution of  $W_j$s to be Weibull that is identical for all individuals. However, by linking the covariates to the lifetime distribution of $W_j$ using the model  in  (\ref{eqxx20}), a greater degree of flexibility is added to the model, as the lifetime of the susceptible is different for each individual depending on the values of covariates. Then, testing the hypothesis $H_0: (\bm \gamma_2', \bm \gamma_3') = \bm 0$ would enable us to infer on the homogeneity of the lifetime distributions across all subjects.\\


\section{Observed data and likelihood functions}\label{sect3}
\hspace{5mm} In clinical studies, right censoring occurs commonly due to patient's discontinuation, duration of study, or lost to follow-up. For this reason, we assume non-informative right censored data in our analysis. In general, for $i=1, \dots, n$, if we consider $Y_i$ to be the actual lifetime and $C_i$ to be the censoring time corresponding to the $i$-th individual, then time to event $T_i$ is defined as  $$T_i=\min\{Y_i, C_i\}.$$ The censoring indicator is given by $\delta_i=I(T_i \le C_i)$ which takes value 1 when the actual lifetime is observed and 0 when the lifetime is right censored at  time $C_i$ for the $i$-th individual. \\

\hspace{5mm} For $i=1, \dots, n$, two sets of covariates $\bm x_i=(x_{i1}, \dots, x_{iq_1})'$ and $\bm z_i=(z_{i1}, \dots, z_{iq_2})'$ are  linked to the parameters $p$ and $\eta$ such that $\eta_i=e^{\bm \alpha' \bm z_i}$ is linked using a log-linear function while $p_i=\frac{e^{\bm \beta' \bm x_i}}{1+e^{\bm \beta' \bm x_i}}$ is linked using a logit function, where $\bm \alpha=(\alpha_1, \dots, \alpha_{q_2})'$ and $\bm \beta=(\beta_0, \beta_1, \dots, \beta_{q_1})'$ are now new model parameters. To circumvent the issue of non-identifiability of parameters associated with DEWP, DLBP or DNB cure models, $\bm \alpha$ is taken without an intercept term and covariate $\bm x_i$ is assumed to be disjoint of $\bm z_i$ in the sense that they have no common elements (see \citealp{li2001identifiability}).   The observed data for $n$ individuals  is then of the form  $(t_i, \delta_i, \bm x_i', \bm z_i')'$,  $i=1, \dots, n$. Thence, the observed data likelihood function can be expressed as
\begin{equation}\label{eqx25}
L(\bm \theta; \bm t, \bm \delta, \bm X, \bm Z) \propto \prod_{i=1}^n f_p(t_i; \bm x_i, \bm z_i, \bm \theta)^{\delta_i}S_p(t_i; \bm x_i, \bm z_i, \bm \theta)^{1-\delta_i},
\end{equation}where $\bm \theta=(\bm \alpha', \bm \beta', \bm \gamma', \phi)'$, $\bm \alpha = (\alpha_1, \dots, \alpha_{q_2})'$, $\bm \beta=(\beta_1, \dots, \beta_{q_1})'$, $\bm \gamma=(\gamma_0, \gamma_1, \bm \gamma'_2, \bm \gamma'_3)'$, $\bm \gamma_2=(\gamma_{21}, \dots, \gamma_{2q_1})'$, $\bm \gamma_3=(\gamma_{31}, \dots, \gamma_{3q_2})'$, $\bm t=(t_1, \dots, t_n)'$, $\bm \delta=(\delta_1, \dots, \delta_n)'$, $\bm X=(\bm x_1, \dots, \bm x_n)$ and $\bm Z=(\bm z_1, \dots, \bm z_n)$. In  (\ref{eqx25}), $f_p(t_i; \bm x_i, \bm z_i, \bm \theta)=f_p(t_i)$ and $S_p(t_i; \bm x_i, \bm z_i, \bm \theta)=S_p(t_i)$ are given by (\ref{eqx8}), (\ref{eqx9}),  (\ref{eqx10}), (\ref{eqx11}), (\ref{eqx17}), (\ref{eqx18}), (\ref{eqx22}), (\ref{eqx23}) and (\ref{eqx24}).\\


\section{Estimation of parameters and standard errors of estimates}\label{sect4}
\hspace{5mm}We implement the EM algorithm for estimating $(\bm \alpha', \bm \beta', \bm \gamma')'$ while $\phi$ is estimated using profile likelihood method. The missing data are introduced by defining indicators $I_i$ that take 0 if the $i$-th individual is cured and 1 otherwise. Note that $I_i=1$ for $i \in \Delta_1$, but $I_i$ is unobserved for $i \in \Delta_0$, with $\Delta_1=\{i: \delta_i=1\}$ and $\Delta_0=\{i: \delta_i=0\}$. For any $i=1, \dots, n$, $\pi_0=\pi_0(\bm \alpha, \bm \beta; \bm x_i, \bm z_i)$ can be obtained from (\ref{eqxx7.5}), (\ref{eqxx9.5}) and (\ref{eqx16}) for the three cure models described in Section \ref{sect2}. Further, let us denote $\bm I=(I_1,\ldots,I_n)'$.  \\

\hspace{5mm}With the complete data denoted by $\{(t_i, \delta_i, \bm x_i', \bm z_i', I_i)',\text{ } i=1, \dots, n\}$, the complete data likelihood function is given by
\begin{equation}
\begin{aligned}\label{eqx26}
& L_c(\bm \theta; \bm t, \bm x, \bm z, \bm \delta, \bm I )   \\ & \propto \prod_{i \in \Delta_1} f_p(t_i; \bm x_i, \bm z_i, \bm \theta) \prod_{i \in \Delta_0} \pi_0(\bm \alpha, \bm \beta; \bm x_i, \bm z_i)^{1-I_i}\{(1-\pi_0(\bm \alpha, \bm \beta; \bm x_i, \bm z_i))S_u(t_i; \bm x_{i}, \bm z_i, \bm \theta)\}^{I_i}\\
\end{aligned}
\end{equation}and the complete data log-likelihood function is then
\begin{equation}
\begin{aligned}
\label{eqx27}
l_c(\bm \theta; \bm t, \bm x, \bm z, \bm \delta, \bm I) =  &\quad \text{constant} + \sum_{i \in \Delta_1} \log f_p(t_i; \bm x_i, \bm z_i, \bm \theta) + \sum_{i \in \Delta_0} (1-I_i)\log \pi_0(\bm \alpha, \bm \beta; \bm x_i, \bm z_i) \\&+\sum_{i \in \Delta_0} I_i\log (1-\pi_0(\bm \alpha, \bm \beta; \bm x_i, \bm z_i)) +\sum_{i \in \Delta_0} I_i \log S_u(t_i; \bm x_{i}, \bm z_i, \bm \theta),
\end{aligned}
\end{equation}
where $S_u(t_i; \bm x_{i}, \bm z_i, \bm \theta)=\frac{S_p(t_i; \bm x_{i}, \bm z_i, \bm \theta)-\pi_0(\bm \alpha, \bm \beta; \bm x_i, \bm z_i)}{1-\pi_0(\bm \alpha, \bm \beta; \bm x_i, \bm z_i)}$ using (\ref{eqx1}). Now, the steps of the EM algorithm proceed as follows.\\

\hspace{5mm}{\it \bf E-step:} For a fixed value $\phi_0$ of $\phi$ and at the $(a+1)$-th iteration of the EM algorithm, we compute the expected value of $l_c(\bm \theta; \bm t, \bm x, \bm z,  \bm \delta, \bm I)$, given the observed data ${\bm O}=\{  (t_i, \delta_i, \bm x_i', \bm z_i', I_{i^*})': i=1,\dots, n,  {i^*} \in \Delta_1\}$ and the current parameter estimates $\hat{\bm \theta}^{*(a)}$ obtained from the $a$-th iteration, where $\bm \theta^{*}=(\bm \alpha', \bm \beta', \bm \gamma')'$. Therefore, from  (\ref{eqx27}), we obtain 
\begin{equation}
\begin{aligned}
\label{eq:4_2}
\mathbb{E}\left(l_c(\bm \theta; \bm t, \bm x, \bm z, \bm \delta, \bm I)\vert \hat{\bm \theta}^{*(a)}, \bm O\right) &= \text{constant} + \sum_{i \in \Delta_1} \log f_p(t_i; \bm x_i, \bm z_i, \bm \theta) + \sum_{i \in \Delta_0} \left(1-\xi_i^{(a)}\right)\log \pi_0(\bm \alpha, \bm \beta; \bm x_i, \bm z_i) \\&+\sum_{i \in \Delta_0} \xi_i^{(a)}\log \left(1-\pi_0(\bm \alpha, \bm \beta; \bm x_i, \bm z_i)\right) +\sum_{i \in \Delta_0} \xi_i^{(a)} \log S_u(t_i; \bm x_{i}, \bm z_i, \bm \theta),
\end{aligned}
\end{equation}where 
\begin{equation}
\begin{aligned}
\xi_i^{(a)}=\mathbb{E}\left(I_i|\hat{\bm \theta}^{*(a)}, {\bm O}\right)=\frac{(1-\pi_0(\bm \alpha, \bm \beta; \bm x_i, \bm z_i))S_u(t_i; \bm x_{i}, \bm z_{i}, \bm \theta)}{S_p(t_i; \bm x_i, \bm z_i, \bm \theta)}\bigg \rvert_{\bm \theta^{*}=\hat{\bm \theta}^{*(a)}}.
\nonumber
\end{aligned}
\end{equation} 
Define $Q\left(\bm \theta^{*}, \bm \xi^{(a)}\right)=\mathbb{E}\left(l_c(\bm \theta; \bm t, \bm x, \bm z, \bm \delta, \bm I ) \vert \hat{\bm \theta}^{*(a)}, \bm O\right)$, where $\bm \xi^{(a)}=\left(\xi_i^{(a)}: i \in \Delta_0\right)'$.\\

\hspace{5mm}{\it \bf M-step:} In the maximization step, we maximize $Q(\bm \theta^{*}, \bm \xi^{(a)})$ with respect to $\bm \theta^{*}$ for finding the ML estimate $\hat{\bm \theta}^{*(a+1)}$ of $\bm \theta^{*}$ at the $(a+1)$-th step of iteration. The numerical maximization is carried out using  Nelder-Mead or Quasi-Newton method, for fixed $\phi_0$. Explicit expressions for $Q(\bm \theta^{*}, \bm \xi^{(a)})$, and the first-order and second-order partial derivatives of $Q(\bm \theta^{*}, \bm \xi^{(a)})$ are presented in Appendices \ref{A3}, \ref{B3} and \ref{C3}, respectively. The iterative process gets terminated at the $(a+1)$-th step if $$ \max_{1 \le k' \le p^*} \left| \frac{ \hat{\theta}_{k'}^{*(a+1)}-{\hat{\theta}_{k'}^{*(a)}}}{ {\hat{ \theta}_{k'}^{*(a)}}}\right| < \epsilon, \text{ }a=1,2, \dots,$$ for some pre-fixed tolerance value of $\epsilon$, where $\hat{\theta}^{*(a')}_{k'}$ is the $k'$-th component of $\hat{\bm \theta}^{*(a')}$ and $p^*$ denotes the dimension of $\bm \theta^*$. \\ 

\hspace{5mm} The estimation of $\phi$ is carried out using the profile likelihood approach since the likelihood function is relatively flat with respect to $\phi$. The {\it E-step} and {\it M-step} are then repeated for all $\phi \in \Phi$, where $\Phi$ denotes the admissible range of $\phi$. The value of $\phi \in \Phi$ that provides the maximum value of the observed log-likelihood function is accepted as the ML estimate $\hat \phi$ of $\phi$. For the DEWP cure model, we made use of the ranges $\Phi=\{-2.0, -1.9, \dots, 2.0\}$, while for the DNB cure model, $\Phi=\{0.10, 0.15, \dots, 7.00\}$.\\

\hspace{5mm}  Under suitable regulatory conditions, it can be established that the ML estimator $\bm {\hat \theta}^*$ of ${\bm \theta}^*$ follows an asymptotic multivariate normal distribution with mean vector $\bm \theta^*$ and covariance matrix $\Sigma(\hat {\bm\theta}^*)$, with an estimate of $\Sigma(\hat {\bm \theta}^*)$ being 
$$ \hat \Sigma(\hat {\bm \theta}^*)= \left\{-\frac{\partial ^2 \log L(\bm \theta; \bm t, \bm \delta, \bm X, \bm Z)}{\partial \bm \theta^* \partial \bm \theta^{*\bm '}}\right\}^{-1}\Bigg \vert_{ {\bm \theta}^*= \hat {\bm \theta}^*}.$$ 
For $\alpha' \in (0,1)$, $100(1-\alpha')\%$ confidence interval (C.I.) of the parameters can then be readily constructed by using the asymptotic normality of the ML estimators.   

\section{Analysis of cutaneous melanoma data}\label{sect5}

\hspace{5mm} For the purpose of illustration of the models and the method of inference developed, we consider the data set `melanoma' available in the \texttt{timereg} package in R. The data set contains information from a historically prospective clinical study in the period 1962-1977 on malignant melanoma with $225$ patients (\citealp{andersen2012statistical}). The following variables are present in the data set: survival time since operation (in years), tumor thickness (in cm), censoring status ($1 \equiv$ died from the disease, $2\equiv$ alive at the end of the study, $3 \equiv$ died from unrelated causes), ulceration status ($1\equiv$ ulcer present, $0\equiv$ ulcer absent), sex ($1\equiv$ male, $0 \equiv$ female), age (in years) and year of operation. Out of $225$ patients, $20$ subjects did not have histological evaluation. Among the remaining $205$ patients, $57$ patients died before the end of $1977$ and the censoring proportion is thus $72.19\%$.\\

\begin{figure}
  \centering
    \includegraphics[width=10cm]{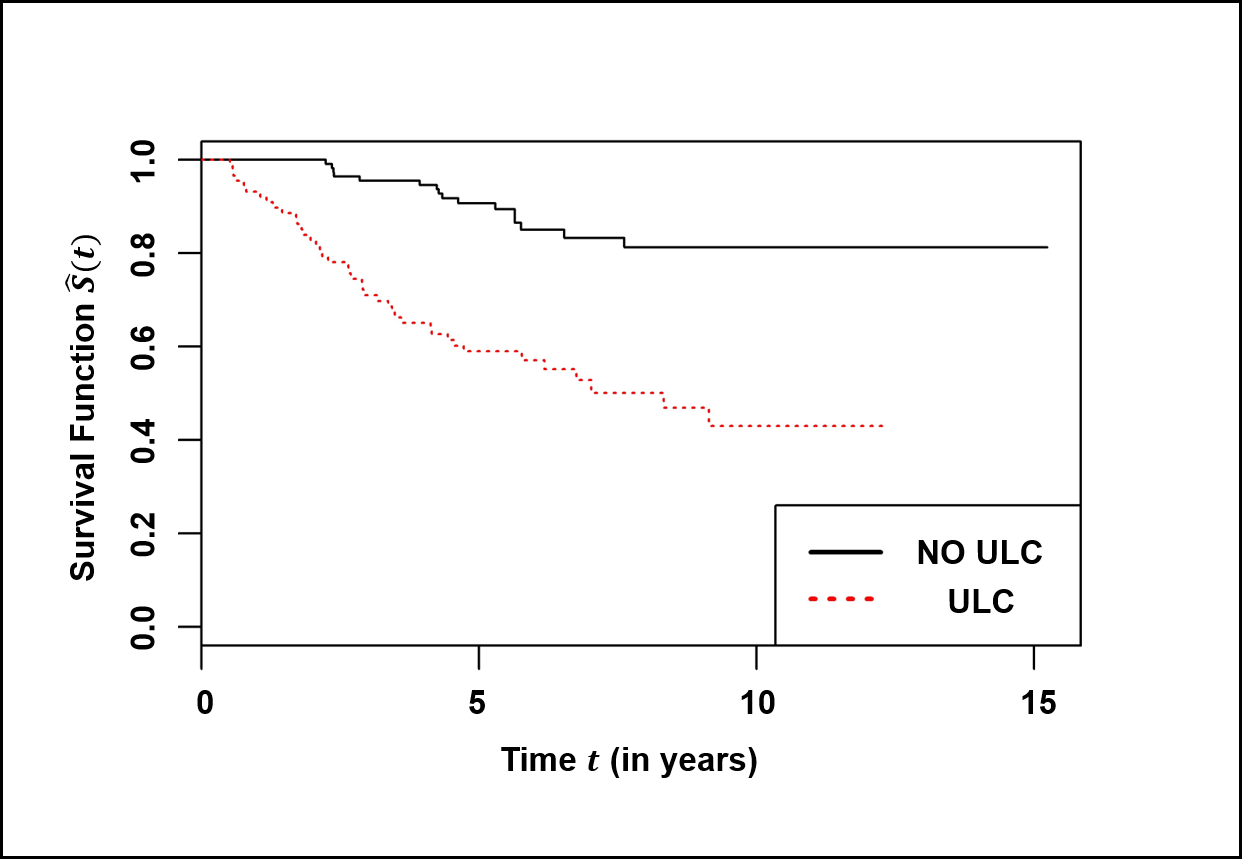}
      \caption{KM plot categorized by the ulceration status}
\label{f:41}
\end{figure}

\hspace{5mm} The observed time (in years) refers to the time since operation till patient's death or the censoring time, with corresponding mean and standard deviation (s.d.) as $5.89$ and $3.07$ years, respectively. For our analysis, ulceration status $z$ (absent: $n=115$; present: $n=90$) and tumor thickness $x$ (in mm) are selected as covariates for the study. $44\%$ of the patients had ulceration status present at the beginning of the study. For this group, mean and s.d. of the tumor thicknesses are  $4.34$ mm and $3.22$ mm, respectively. For the  group with ulceration status as absent, the mean and s.d. of the tumor thicknesses  are $1.81$ mm and $2.19$ mm, respectively.  The histograms of the tumor thickness for both the groups show positively skewed distributions. Figure \ref{f:41} represents the Kaplan-Meier (KM) plot categorized by the ulceration status, which possibly indicates the presence of cure proportion in the data. \\

\hspace{5mm}  Note that  $\phi=0$ reduces DEWP cure model to the usual destructive Poisson (DP) cure model, studied originally by \cite{rodrigues2011destructive}. Similarly, we obtain exponentially weighted Poisson (EWP) and Poisson cure models by setting $p=1$ and $(p=1, \phi=0)$, respectively.   Destructive geometric (DG), negative binomial (NB) and geometric cure models are obtained from the DNB cure model by considering $\phi=1$, $p=1$ and $(\phi=1, p=1)$, respectively. $p=1$ represents cases wherein no destructive mechanism of the malignant cells is involved. When $p=1$, we link both the covariates to $\eta$ using log-linear link function of the form $\eta=\exp(\beta_0+ \beta_1  x + \alpha z)$. \\

\hspace{5mm} In Table \ref{T:41}, the number of parameters fitted ($\tilde k$), maximized log-likelihood ($\hat l$) values, Akaike's Information Criterion (AIC) and Bayesian Information Criterion (BIC) values for all fitted models, including all sub-models, are presented. For $\hat \phi=5.2$, the DNB cure model provides the best fit to the data with highest maximized log-likelihood ($-199.108$) and minimum AIC ($414.216$) values.  The estimate, standard error (SE), lower confidence limit (LCL) and upper confidence limit (UCL) of all parameters for the three main models are presented in Table \ref{T:42}. For validating heterogeneity among the lifetimes of susceptible individuals, we  carry out a test of hypothesis $H_0: \gamma_2=\gamma_3=0$ vs. $H_1: \text{at least one inequality in } H_0$ for the DNB model with $\phi=5.2$. The resultant $p$-value of the test is $0.061$, indicating rejection of $H_0$ at 10\% level, and the corresponding maximized log-likelihood value for the reduced model is $-201.908$.  Moreover, on testing $H_0:\phi=0$ for the full DNB model, we found the $p$-value to be $0.027$ which does reveal  that DNB models provide a better fit than the DG model for this data. It can be observed  from Table \ref{T:41} that incorporating destructive mechanism to the cure models have resulted in better log-likelihood, AIC and BIC values, which justifies the practicality of applying destructive cure models over regular cure models. \\

\begin{table}[htbp]
\caption{Maximized log-likelihood ($\hat l$), AIC and BIC values for some destructive cure models}
\centering
\begin{tabular*}{\textwidth}{l @{\extracolsep{\fill}} cccc}    \hline
    { Fitted Model \qquad} &   ${ {\tilde k}} $ & $ \hat { l}$ & { AIC} & { BIC}\\
\hline
     DEWP ($\hat \phi = -0.7$) \qquad &  8 & -202.253 &  420.506 & 447.090\\

     \quad DP  & 7 & -203.433 & 420.865 & 444.126 \\
\quad EWP ($\hat\phi=-1.5$) & 8 & -205.054 & 426.108 & 452.693 \\
\quad Poisson & 7 &-205.054 & 424.108 & 447.370 \\
\hline
 DLBP &  7 & -204.979 & 423.959 & 447.220 \\
 \hline   
 DNB ($\hat \phi=5.2$) &  8 & {\bf -199.108} & {\bf 414.216} & 440.800\\
     \quad DG &7 & -201.536 & 417.073 & {\bf 440.334}\\

        \quad NB ($\hat \phi = 6.9$)      &8    & -199.973 & 415.946 & 442.531\\
   \quad Geometric  &7 & -204.027 & 422.053 & 445.314\\
    \hline
 \end{tabular*}
\label{T:41}
\end{table}
\begin{table}[htbp]
\caption{ Estimate, SE,  LCL and UCL for DEWP, DLBP and DNB cure models for the cutaneous melanoma data}
\begin{tabular*}{\textwidth}{   c @{\extracolsep{\fill}}    c  c  c  c  c  c  c  c  c  }
\hline
{Fitted Model} &   { Measure} & $\alpha$ & $\beta_0$ & $\beta_1$ & $\gamma_0$ &  $\gamma_1$ &  $\gamma_2$ &  $\gamma_3$ & $\phi$ \\
\hline

 & EST & 0.761 & -1.985 &  1.265 &  1.845 &  7.423 &  0.112 &  0.305 &  -0.7 \\

DEWP   & SE & 0.218 &0.909  & 0.646 & 0.219 & 1.904 & 0.043 & 0.492 & -\\

  & LCL & 0.333 & -3.768 & -0.002 &  1.414 &  3.689 & 0.027 &-0.660 & - \\

 & UCL & 1.188 & -0.202 &  2.532 &  2.276 & 11.156 &  0.196 & 1.270 & -\\
\hline
 &  EST & 1.527 & -2.119 & 0.081 &  1.822 &  8.011 &  0.115 &  0.433  & - \\
 
  DLBP & SE & 0.529 & 0.454 & 0.053 & 0.224 & 2.723 & 0.046 & 0.611 & - \\

 & LCL & 0.489 & -3.009 & -0.023 &  1.382 &  2.672 &  0.024 & -0.765 & - \\

  & UCL & 2.565 & -1.229 &  0.186 &  2.263 & 13.349 &  0.207 &  1.633 & - \\
\hline

 & EST  & 3.670  & -2.602 &   1.081 &  2.845 &  7.282 &  0.192 & -1.596 & 5.2 \\
  DNB     & SE & 1.205 & 0.925 & 0.537 & 0.328 & 1.342 & 0.071 & 1.236 & - \\

 & LCL & 1.306 & -4.416 &  0.027 &  2.201 &  4.650 &  0.052 & -4.019 & - \\

 & UCL & 6.033 &  -0.788 & 2.136 &  3.489 &  9.913 &  0.332 &  0.826 & - \\
\hline 
\end{tabular*}
\label{T:42}
\end{table}
\hspace{5mm} Table \ref{T:43} demonstrates the effects of using different link functions (L1-L4) on  maximized log-likelihood value for  the main three destructive cure models. Considering all four possible combinations, we found link L1 (see Section \ref{sect3}) provided large  $\hat l$ consistently. Because the DNB cure model with $\hat \phi=5.2$ provided the best fit with link L1, we use this link for all our subsequent analyses.\\
\begin{table}[htbp]
\caption{Maximized log-likelihood values for destructive cure models with various link functions}
\begin{tabular*}{\textwidth}{ l @{\extracolsep{\fill}}c c c  }
    \hline
     Link Function &   Model  & $\hat \phi$ & $ \hat { l}$ \\
    \hline
    \multirow{3}{*}{L1: $\eta=e^{\alpha z}, \frac{e^{\beta_0+\beta_1 x}}{1+e^{\beta_0+\beta_1 x}}^{**}$ }   & DEWP &-0.7  & -205.253\\
                                                                                                                                                                       & DLBP & - & -204.979  \\
    															       & DNB & 5.2 & -199.108\\

    \hline
    \multirow{3}{*}{L2: $\eta=e^{\alpha x}, \frac{e^{\beta_0+\beta_1 z}}{1+e^{\beta_0+\beta_1 z}}$} & DEWP & -0.4 & -205.055 \\
                                                                                                                                                                       & DLBP & - &-208.289  \\
    															       & DNB & 6.9 & -199.962\\
\hline

   \multirow{3}{*}{L3: $\eta=e^{\alpha_0 + \alpha_1 z}, \frac{e^{\beta x}}{1+e^{\beta x}}$ } & DEWP & -1.0 & -203.994\\
                                                                                                                                                                       & DLBP & - & -206.786  \\
    															       & DNB & 7.2 & -201.085\\

\hline

   \multirow{3}{*}{L4: $\eta=e^{\alpha_0 + \alpha_1 x}, \frac{e^{\beta z}}{1+e^{\beta z}}$ } & DEWP & -0.2 & -205.302 \\
                                                                                                                                                                       & DLBP & - & -206.667 \\
    															       & DNB & 6.4 & -200.313\\
\hline
\end{tabular*}
\raggedright{ { \footnotesize $^{**}$This link is used for subsequent analyses.}}
\label{T:43}
\end{table}

\begin{figure}
    \centering
    \begin{subfigure}[b]{0.45\textwidth}
       \includegraphics[width=\textwidth]{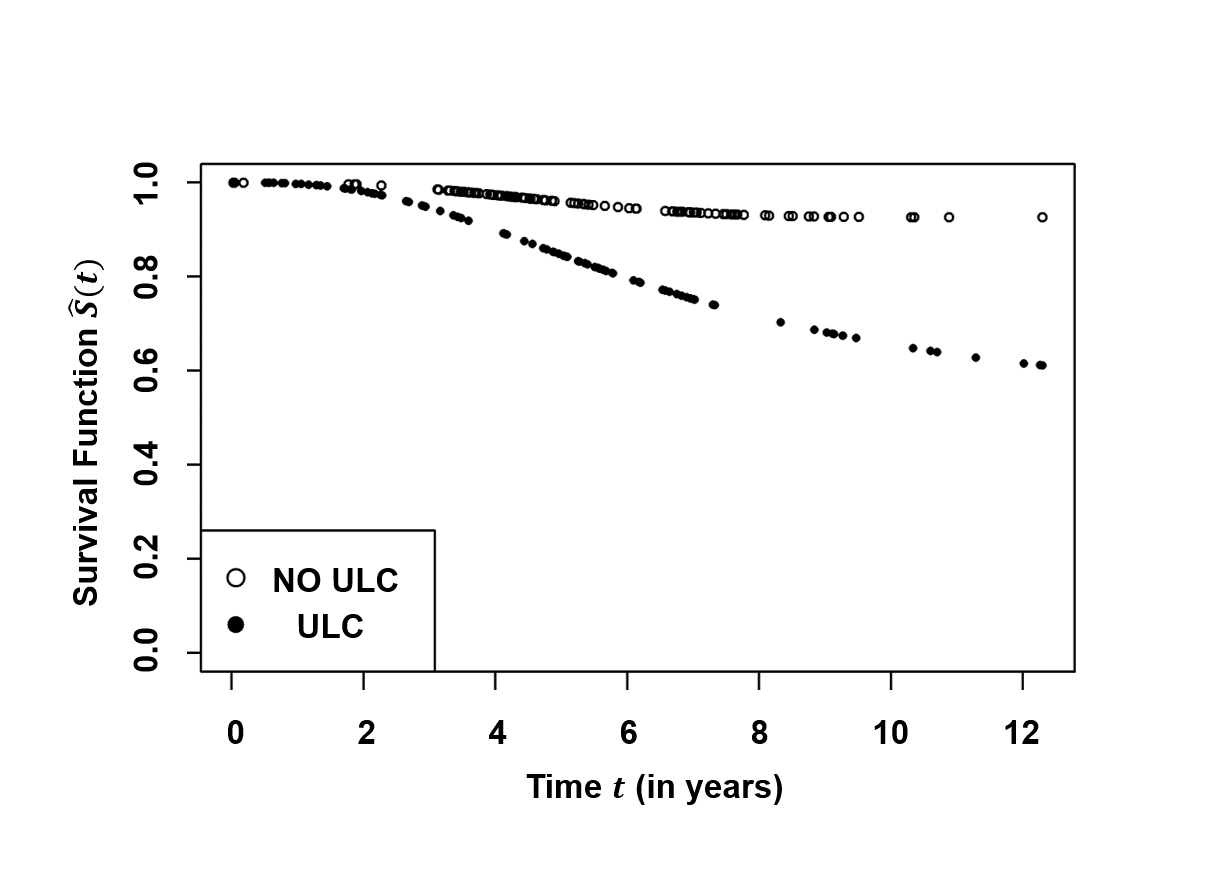}
        \caption{Survival plots stratified by the ulceration status with tumor thickness = 0.320 mm}
        \label{f:42a}
    \end{subfigure}
\quad
   \begin{subfigure}[b]{0.45\textwidth}
      \includegraphics[width=\textwidth]{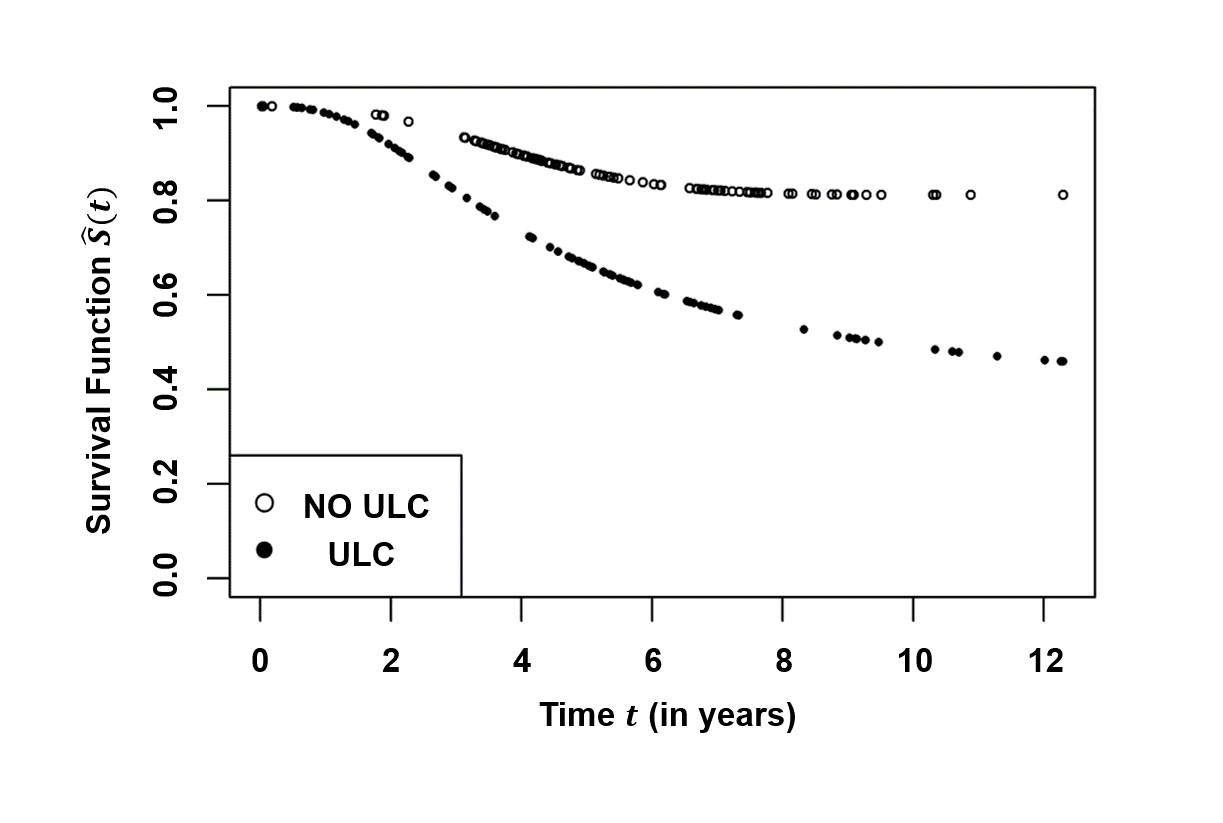}
        \caption{Survival plots stratified by the ulceration status with tumor thickness = 1.940 mm}
        \label{f:42b}
    \end{subfigure}
    \begin{subfigure}[b]{0.55\textwidth}
        \includegraphics[width=\textwidth]{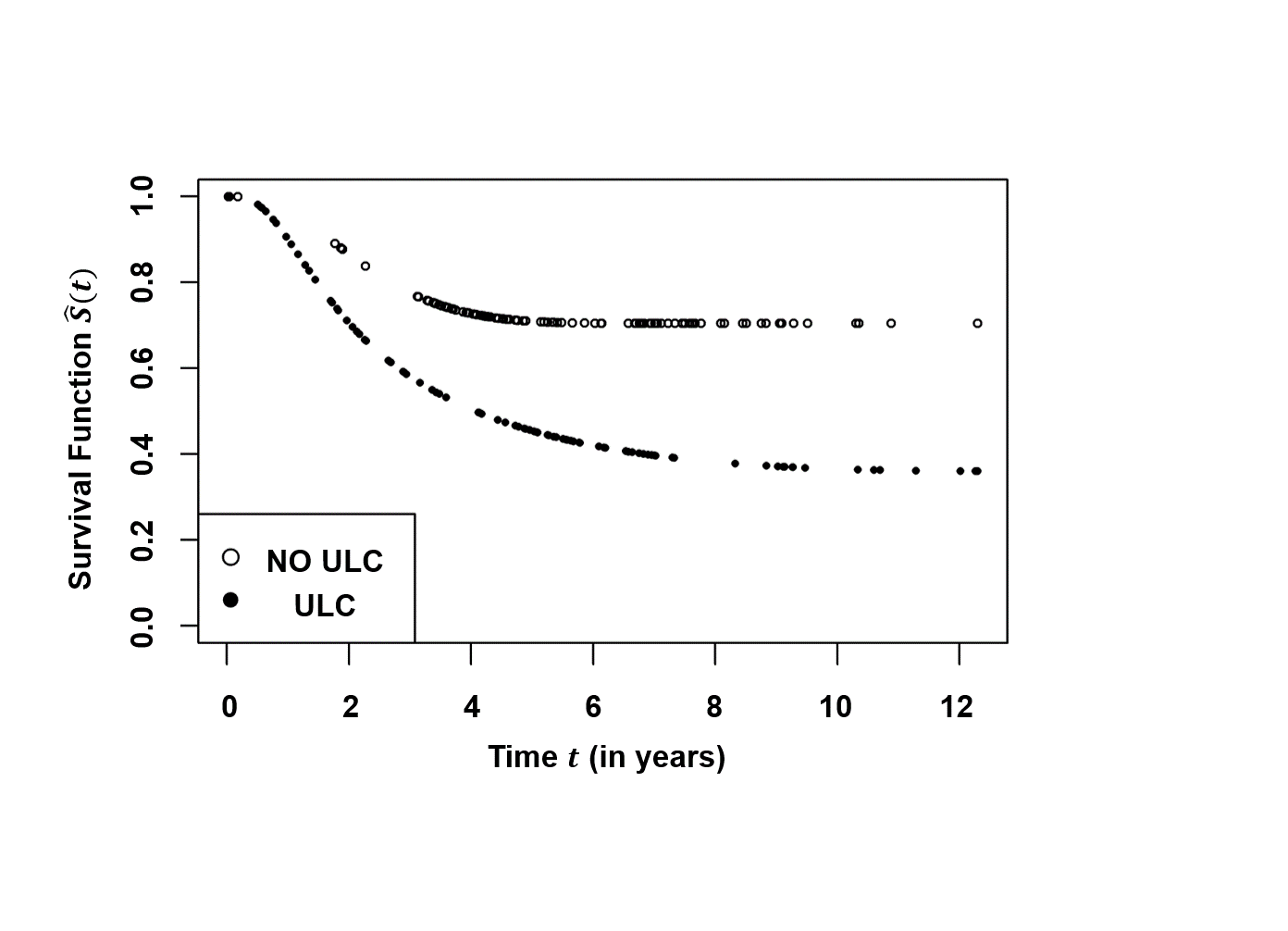}
        \caption{Survival plots stratified by the ulceration status with tumor thickness = 8.320 mm}
        \label{f:42c}
    \end{subfigure}
    \caption{Survival plots stratified by the ulceration status for different thickness levels}\label{f:42}
\end{figure}

\begin{figure}
  \centering
    \includegraphics[width=10cm]{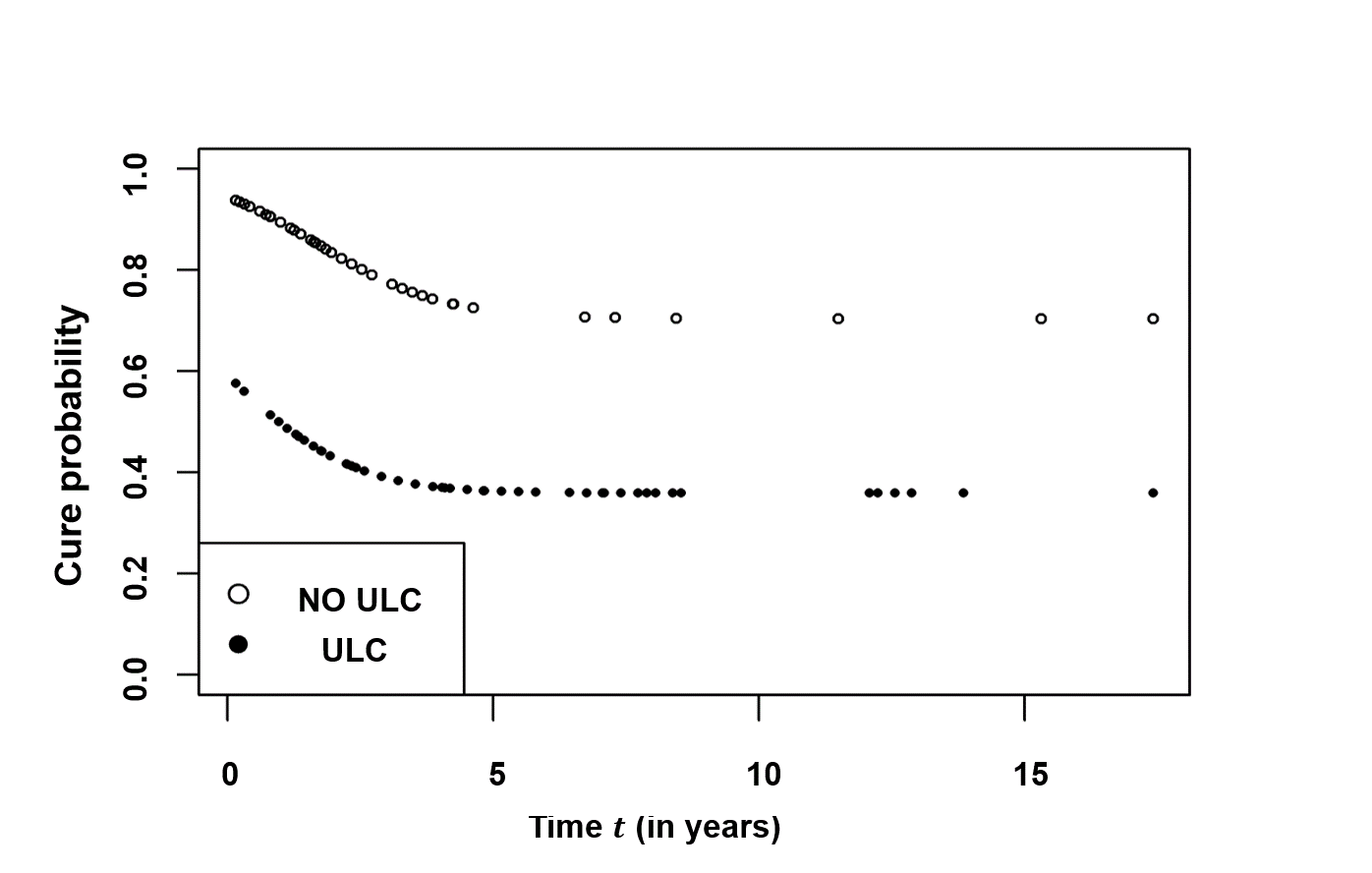}
      \caption{Cure probability vs. tumor thickness stratified by the ulceration status}
\label{f:43}
\end{figure}

\hspace{5mm} We choose three representative values of the tumor thickness: $0.320$, $1.940$ and $8.320$ mm corresponding to the $5$-th, $50$-th and $95$-th percentiles, and plot corresponding long-term survival functions stratified by the ulceration status (Figure \ref{f:42a}-\ref{f:42c}). The estimated survival function values are found to be higher for the group with ulceration status as absent and smaller tumor thickness values. Figure \ref{f:43} represents the estimated cure probabilities against tumor thickness values stratified by the ulceration status. A non-parametric test of difference suggests a significant difference ($p$-value  $<2.2$ x $10^{-16}$) between cure probabilities of the two ulcer groups.

\section{Simulation study}\label{sect6}
\hspace{5mm} We now assess the performance of the proposed method of estimation and inference by using an extensive Monte Carlo simulation study. We generate data set in a way that mimics the real data on cutaneous melanoma discussed in the last section.  For this purpose, we define a random variable $U$, where $U \sim$ Uniform $(0,1)$. If $U \le 0.44$, we assign a  r.v. $Z = 1$, otherwise $Z=0$, where $Z$ denotes the ulceration status for each subject. For simulating the tumor thickness data, we plot histograms of tumor thickness ($X$) values of individuals from the cutaneous melanoma study. The histograms reveal positively skewed curves for both  ulceration statuses; the means and standard deviations  are as given in Section \ref{sect5}. Thus, for $Z=1$, we assume $X$ to follow Weibull distribution with  shape and scale parameters as $\alpha_1$ and $\alpha_2$, respectively. $\alpha_1$ and $\alpha_2$ are estimated by the method of moments by  equating $\alpha_2 \Gamma(1+1/\alpha_1)$ to 4.34 and $ \alpha_2^2\left[\Gamma\left(1+\frac{2}{\alpha_1}\right) - \left(\Gamma\left(1+\frac{1}{\alpha_1}\right)\right)^2\right]$ to $(3.22)^2$. We generate $X$ using the  values of $\alpha_1$ and $\alpha_2$ determined this way. A similar approach is taken to generate $X$ for $Z=0$, where we assume $X$ to follow Weibull ($\alpha_3$, $\alpha_4$) with $\alpha_3$ and $\alpha_4$  being estimated from $\alpha_4 \Gamma(1+1/\alpha_3) = 1.81$ and $ \alpha_4^2\left[\Gamma\left(1+\frac{2}{\alpha_3}\right) - \left(\Gamma\left(1+\frac{1}{\alpha_3}\right)\right)^2\right] = (2.19)^2$. As mentioned before, we linked $\eta$ to $z$ using $\eta=e^{\alpha z}$ and $p$ to $x$ using $p=\frac{e^{ \beta_0 +  \beta_1 x}}{1+e^{ \beta_0 +  \beta_1 x}}$, wherein an intercept term is not taken in the link for  $\eta$  to  avoid non-identifiability. Note that $\eta=1$ whenever $z=0$. Also, a higher value of $\eta$ signifies greater number of initial competing causes ($M$). So, we assume $\eta$ to be more than 1 for $z=1$ since patients with the ulceration status as  `present' are likely to have greater values of $M$. Following the work of \cite{pal2016likelihood}, we assume $\eta=3$ for $z=1$; thereby, we obtain the true value of $\alpha = 1.099$.  In order to determine true values of $\beta_0$ and $\beta_1$, we use $x_{\min} = \min \{x\}=0.1$ mm and $x_{\max} = \max \{x\}=17.42$ mm. As the link $p=\frac{e^{ \beta_0 +  \beta_1 x}}{1+e^{ \beta_0 +  \beta_1 x}}$ is  monotonically increasing in $x$,  we set $p_{\min} =\min \{p\}$ and $p_{\max} = \max \{p\}$ and link them to $x_{\min}$ and $x_{\max}$, respectively.  Two such choices of ($p_{\min}$, $p_{\max}$) are considered, {\it viz.}, (0.2, 0.6) and (0.3, 0.9), representing scenarios of lower and higher proportions of active number of competing causes.  The true values of $\beta_0$ and $\beta_1$ change depending on the generated values of $x$ for each simulation.  \\

\hspace{5mm}$M$ is generated from weighted Poisson distribution with parameter  $\eta$. For exponentially weighted Poisson cure model, we set   $\phi=0.2$ and $-0.5$, and for negative binomial cure model, we take $\phi=0.5$ and $5.2$. For the length-biased Poisson, $M$ is generated from  Poisson ($\eta$) + 1 distribution. Given $M=m > 0$, $D$ is generated from binomial distribution with success probability $p$ and $m$ as the number of trials. If $M=0$, we set $D=0$.  The true values of the lifetime parameters $(\gamma_0, \gamma_1, \gamma_2, \gamma_3)'$ are set to be $(1.657, 3.764, -0.005, 0.023)'$, which are indeed the parameter estimates obtained from the real data. If $D=d >0$,  we generate $W_1, \ldots, W_d$ from Weibull distribution with  shape  $\gamma_0$ and scale $\gamma_1 \exp \left(-\frac{\gamma_2 x +\gamma_3 z}{\gamma_0}\right)$. We then define lifetime $Y=\min\{W_1, \ldots, W_d\}$ and the censoring time $C$ is assumed to be distributed exponentially with rate parameter $\psi$. Hence, the observed time $T$ is defined as $T=\min\{Y, C\}$. Again, if $D=0$, we simply assign $T=C$. To assess the effect of censoring on the proposed methodology, we consider three different scenarios: $\psi = 0.05, 0.15$ and $0.25$ representing low, medium and high censoring levels, respectively. On examining $\psi \in \{0.01, 0.02, \ldots, 1.50\}$ and comparing the proportion of censoring (i.e., no. of times $Y>C$) in 1000 replications, we observed that  $\psi=0.05, 0.15$ and $0.25$ correspond to 52$\%$, 64$\%$ and 72$\%$ of censoring percentages, respectively. $\psi$ as low as 0.01 gives 45$\%$ of censoring whereas $\psi=1.50$ results in 95$\%$ of censored observations. We took the sample size as $n=400$, though similar results were produced for some other choices of $n$, but are not presented here for brevity.\\



\begin{table}[htbp]
\caption{EST, SE, bias, RMSE, 95$\%$ CI and CP for the destructive exponentially weighted Poisson cure model with $\phi=-0.5$ based on $n=400$}
\begin{tabular*}{\textwidth}{c @{\extracolsep{\fill}}  c c c c c c c c c }
\hline
     ($p_{\min}, p_{\max}$) & $\psi$ & $\theta$ &     True Value &       EST &         SE &      BIAS &       RMSE &         95$\%$ CI &         CP  \\
\hline
       (0.2, 0.6)& 0.05& $\alpha$ &      1.099 &      1.095 &      0.268 &     -0.004 &      0.375 & (0.570, 1.620) &      0.913 \\

&&$\beta_0$ &     -1.387 &     -1.472 &      0.298 &     -0.086 &      0.616 & (-2.056, -0.889) &      0.598 \\

&&$\beta_1$ &      0.080 &      0.145 &      0.091 &      0.041 &      0.129 & (-0.032, 0.323) &      0.956 \\

&&$\gamma_0$ &      1.658 &      1.777 &      0.153 &      0.120 &      0.228 & (1.478, 2.077) &      0.906 \\

&&$\gamma_1$ &      3.765 &      3.822 &      0.526 &      0.057 &      0.716 & (2.790, 4.853) &      0.933 \\

&&$\gamma_2$ &     -0.005 &     -0.017 &      0.042 &     -0.012 &      0.057 & (-0.098, 0.064) &      0.941 \\

&&$\gamma_3$ &      0.024 &     -0.066 &      0.297 &     -0.090 &      0.409 & (-0.649, 0.517) &      0.947 \\

   &&$\phi$ &     -0.500 &     -0.466 &          - &          - &          - &          - &          - \\
\hline

 (0.3, 0.9)& 0.05& $\alpha$ &      1.099 &      1.088 &      0.210 &     -0.011 &      0.293 & (0.677, 1.499) &      0.932 \\

&&$\beta_0$ &     -0.849 &     -1.039 &      0.308 &     -0.191 &      0.635 & (-1.641, -0.436) &      0.548 \\

&&$\beta_1$ &      0.163 &      0.279 &      0.164 &      0.102 &      0.242 & (-0.042, 0.599) &      0.891 \\

&&$\gamma_0$ &      1.658 &      1.797 &      0.128 &      0.139 &      0.211 & (1.546, 2.047) &      0.833 \\

&&$\gamma_1$ &      3.765 &      3.839 &      0.430 &      0.074 &      0.598 & (2.995, 4.682) &      0.944 \\

&&$\gamma_2$ &     -0.005 &     -0.025 &      0.036 &     -0.019 &      0.052 & (-0.096, 0.047) &      0.929 \\

&&$\gamma_3$ &      0.024 &     -0.137 &      0.248 &     -0.161 &      0.366 & (-0.623, 0.348) &      0.897 \\

 &&$\phi$ &     -0.500 &     -0.406 &          - &          - &          - &          - &          - \\
\hline

 (0.2, 0.6)& 0.15& $\alpha$ &       1.099 &      1.093 &      0.329 &     -0.006 &      0.447 & (0.449, 1.738) &      0.942 \\

&&$\beta_0$ &     -1.386 &     -1.515 &      0.365 &     -0.129 &      0.636 & (-2.231, -0.799) &      0.755 \\

&&$\beta_1$ &      0.103 &      0.155 &      0.112 &      0.052 &      0.161 & (-0.065, 0.375) &      0.952 \\

&&$\gamma_0$ &      1.658 &      1.790 &      0.184 &      0.132 &      0.275 & (1.429, 2.151) &      0.911 \\

&&$\gamma_1$ &      3.765 &      3.789 &      0.721 &      0.024 &      0.989 & (2.375, 5.203) &      0.907 \\

&&$\gamma_2$ &     -0.005 &     -0.018 &      0.058 &     -0.013 &      0.080 & (-0.132, 0.096) &      0.941 \\

&&$\gamma_3$ &      0.024 &     -0.087 &      0.410 &     -0.111 &      0.572 & (-0.892, 0.717) &      0.928 \\

   &&$\phi$ &     -0.500 &     -0.451 &          - &          - &          - &          - &          - \\
\hline

 (0.3, 0.9)& 0.15& $\alpha$ &       1.099 &      1.079 &      0.258 &      -0.02 &      0.355 & (0.574, 1.584) &      0.927 \\

&&$\beta_0$ &     -0.848 &     -1.026 &      0.399 &     -0.178 &      0.683 & (-1.807, -0.245) &      0.729 \\

&&$\beta_1$ &      0.197 &      0.385 &      0.281 &      0.210 &      0.406 & (-0.166, 0.937) &      0.932 \\

&&$\gamma_0$ &      1.658 &      1.795 &      0.153 &      0.137 &      0.238 & (1.496, 2.094) &      0.877 \\

&&$\gamma_1$ &      3.765 &      3.830 &      0.595 &      0.066 &      0.812 & (2.665, 4.996) &      0.934 \\

&&$\gamma_2$ &     -0.005 &     -0.029 &      0.049 &     -0.023 &      0.069 & (-0.124, 0.067) &      0.911 \\

&&$\gamma_3$ &      0.024 &     -0.135 &      0.338 &     -0.159 &      0.485 & (-0.797, 0.527) &      0.916 \\

   &&$\phi$ &     -0.500 &     -0.435 &          - &          - &          - &          - &          - \\
\hline

 (0.2, 0.6)& 0.25& $\alpha$ &      1.099 &      1.072 &      0.413 &     -0.027 &      0.563 & (0.262, 1.882) &      0.934 \\

&&$\beta_0$ &     -1.387 &     -1.489 &      0.470 &     -0.103 &      0.732 & (-2.410, -0.569) &      0.841 \\

&&$\beta_1$ &      0.109 &      0.170 &      0.145 &      0.066 &      0.208 & (-0.115, 0.455) &      0.961 \\

&&$\gamma_0$ &      1.658 &      1.817 &      0.219 &      0.159 &      0.328 & (1.387, 2.246) &      0.903 \\

&&$\gamma_1$ &      3.765 &      3.789 &      0.990 &      0.025 &      1.332 & (1.850, 5.729) &      0.912 \\

&&$\gamma_2$ &     -0.005 &     -0.017 &      0.077 &     -0.011 &      0.105 & (-0.168, 0.135) &      0.936 \\

&&$\gamma_3$ &      0.024 &     -0.115 &      0.552 &     -0.139 &      0.754 & (-1.198, 0.968) &      0.936 \\

   &&$\phi$ &     -0.500 &     -0.455 &          - &          - &          - &          - &          - \\
\hline
 (0.3, 0.9)& 0.25& $\alpha$ &      1.099 &      1.115 &      0.335 &      0.016 &      0.453 & (0.459, 1.771) &      0.936 \\

&&$\beta_0$ &     -0.847 &     -1.037 &      0.483 &     -0.190 &      0.769 & (-1.983, -0.091) &      0.839 \\

&&$\beta_1$ &      0.148 &      0.363 &      0.282 &      0.186 &      0.407 & (-0.190, 0.915) &      0.911 \\

&&$\gamma_0$ &      1.658 &      1.792 &      0.179 &      0.135 &      0.266 & (1.441, 2.143) &      0.911 \\

&&$\gamma_1$ &      3.765 &      3.820 &      0.821 &      0.056 &      1.097 & (2.212, 5.429) &      0.927 \\

&&$\gamma_2$ &     -0.005 &     -0.023 &      0.062 &     -0.018 &      0.089 & (-0.145, 0.099) &      0.907 \\

&&$\gamma_3$ &      0.024 &     -0.190 &      0.460 &     -0.214 &      0.642 & (-1.09, 0.711) &      0.928 \\

   &&$\phi$ &     -0.500 &     -0.439 &          - &          - &          - &          - &          - \\
\hline

\end{tabular*}  
\label{T:4}
\footnote*{\footnotesize{EST: Parameter Estimate, SE: Standard Error, BIAS: Bias in Estimation, RMSE: Root Mean Squared Error, CI: Confidence Interval, CP: Coverage Probability (Nominal Level of 95\%)} }
\end{table}


\begin{table}[h!]
\caption{EST, SE, bias, RMSE, 95$\%$ CI and CP for the destructive length-biased Poisson cure model  based on $n=400$}
\begin{tabular*}{\textwidth}{c @{\extracolsep{\fill}}  c c c c c c c c c }
\hline
     ($p_{\min}, p_{\max}$) & $\psi$ & $\theta$ &     True Value &       EST &         SE &      BIAS &       RMSE &         95$\%$ CI &         CP  \\
\hline

        (0.2, 0.6) & 0.05 &  $\alpha$ &      1.099 &       1.060 &      0.252 &     -0.039 &      0.341 &      (0.567,      1.553) &      0.952 \\

       &&     $\beta_0$ &      -1.386 &     -1.392 &      0.163 &     -0.005 &       0.220 &     (-1.711,     -1.072) &      0.957 \\

       &&    $\beta_1$ &      0.085 &      0.108 &      0.044 &      0.005 &      0.059 &      (0.022,      0.194) &      0.954 \\

       &&    $\gamma_0$ &       1.658 &      1.792 &      0.106 &      0.135 &      0.186 &      (1.584,          2.000) &      0.774 \\

       &&    $\gamma_1$ &       3.765 &      3.922 &      0.326 &      0.157 &      0.466 &      (3.284,       4.560) &      0.926 \\

       &&    $\gamma_2$ &      -0.005 &     -0.028 &      0.033 &     -0.023 &      0.048 &     (-0.092,      0.036) &      0.892 \\

       &&    $\gamma_3$ &       0.024 &     -0.139 &      0.207 &     -0.163 &      0.316 &     (-0.544,      0.266) &      0.874 \\
\hline


 (0.3, 0.9) & 0.05 &  $\alpha$ &      1.099 &      1.052 &      0.247 &     -0.047 &      0.333 &      (0.568,      1.536) &      0.961 \\

       &&     $\beta_0$ &      -0.847 &     -0.852 &      0.168 &     -0.004 &      0.233 &     (-1.181,     -0.522) &      0.942 \\

       &&    $\beta_1$ &      0.205 &      0.186 &      0.064 &      0.012 &      0.087 &      (0.062,      0.311) &      0.942 \\

       &&    $\gamma_0$ &       1.658 &      1.811 &      0.093 &      0.153 &      0.187 &      (1.627,      1.994) &      0.653 \\

       &&    $\gamma_1$ &       3.765 &      3.995 &      0.287 &      0.231 &      0.438 &      (3.433,      4.558) &      0.904 \\

       &&    $\gamma_2$ &      -0.005 &     -0.048 &       0.03 &     -0.043 &      0.056 &     (-0.108,      0.011) &      0.711 \\

       &&    $\gamma_3$ &       0.024 &     -0.254 &      0.203 &     -0.278 &      0.371 &    (-0.652,      0.143) &      0.736 \\
\hline


        (0.2, 0.6) & 0.15 &  $\alpha$ &      1.099 &      1.062 &      0.333 &     -0.037 &      0.435 &      (0.409,      1.714) &      0.975 \\

       &&     $\beta_0$ &      -1.386 &     -1.399 &      0.208 &     -0.013 &      0.273 &     (-1.807,     -0.992) &      0.964 \\

       &&    $\beta_1$ &      0.097 &      0.109 &       0.060 &      0.006 &      0.078 &     (-0.009,      0.227) &      0.966 \\

       &&    $\gamma_0$ &       1.658 &      1.784 &      0.126 &      0.126 &      0.203 &      (1.537,      2.031) &      0.848 \\

       &&    $\gamma_1$ &       3.765 &      3.895 &      0.448 &       0.130 &      0.599 &      (3.017,      4.773) &      0.963 \\

       &&    $\gamma_2$ &      -0.005 &     -0.027 &      0.046 &     -0.022 &      0.064 &     (-0.116,      0.062) &       0.920 \\

       &&    $\gamma_3$ &       0.024 &     -0.165 &      0.291 &     -0.189 &      0.422 &     (-0.736,      0.407) &      0.894 \\
\hline


        (0.3, 0.9) & 0.15 &  $\alpha$ &      1.099 &      1.045 &      0.346 &     -0.054 &      0.447 &      (0.367,      1.724) &       0.980 \\

       &&     $\beta_0$ &      -0.849 &     -0.852 &      0.216 &     -0.004 &      0.286 &    (-1.276,     -0.428) &      0.956 \\

       &&    $\beta_1$ &      0.165 &       0.180 &      0.084 &      0.004 &       0.110 &      (0.016,      0.344) &      0.939 \\

       &&    $\gamma_0$ &       1.658 &      1.806 &      0.111 &      0.148 &        0.200 &      (1.589,      2.023) &      0.748 \\

       &&    $\gamma_1$ &       3.765 &      3.965 &      0.391 &        0.200 &      0.537 &      (3.198,      4.732) &      0.957 \\

       &&    $\gamma_2$ &      -0.005 &     -0.045 &       0.040 &      -0.040 &      0.065 &     (-0.124,      0.034) &      0.827 \\

       &&    $\gamma_3$ &       0.024 &     -0.283 &      0.295 &     -0.307 &      0.475 &     (-0.861,      0.294) &      0.856 \\
\hline


        (0.2, 0.6) & 0.25 &  $\alpha$ &      1.099 &      1.055 &      0.455 &     -0.044 &      0.566 &      (0.163,      1.947) &      0.986 \\

       &&     $\beta_0$ &      -1.387 &     -1.406 &       0.270 &      -0.02 &      0.341 &     (-1.936,     -0.876) &      0.981 \\

       &&    $\beta_1$ &      0.093 &      0.108 &      0.081 &      0.005 &      0.102 &      (-0.050,      0.266) &      0.962 \\

       &&    $\gamma_0$ &       1.658 &      1.784 &      0.147 &      0.127 &      0.225 &      (1.496,      2.072) &      0.892 \\

       &&    $\gamma_1$ &       3.765 &      3.892 &      0.612 &      0.127 &      0.791 &      (2.692,      5.091) &      0.973 \\

       &&    $\gamma_2$ &      -0.005 &     -0.029 &      0.062 &     -0.023 &      0.082 &      (-0.150,      0.092) &       0.950 \\

       &&    $\gamma_3$ &       0.024 &      -0.150 &      0.408 &     -0.174 &      0.539 &     (-0.949,      0.649) &      0.961 \\

\hline

       (0.3, 0.9) & 0.25 &  $\alpha$ &      1.099 &      1.031 &      0.553 &     -0.068 &      0.659 &     (-0.053,      2.115) &      0.993 \\

       &&     $\beta_0$ &      -0.847 &     -0.839 &      0.284 &      0.009 &      0.356 &     (-1.396,     -0.282) &      0.979 \\

       &&    $\beta_1$ &      0.165 &      0.176 &      0.108 &          0.000 &      0.138 &     (-0.037,      0.388) &      0.947 \\

       &&    $\gamma_0$ &       1.658 &      1.799 &      0.128 &      0.141 &      0.212 &      (1.548,       2.050) &       0.830 \\

       &&    $\gamma_1$ &       3.765 &      3.995 &      0.547 &       0.230 &      0.716 &      (2.923,      5.067) &      0.978 \\

       &&    $\gamma_2$ &      -0.005 &     -0.042 &      0.052 &     -0.037 &      0.076 &     (-0.143,      0.059) &      0.885 \\

       &&    $\gamma_3$ &       0.024 &     -0.292 &      0.421 &     -0.316 &      0.593 &     (-1.117,      0.534) &      0.939 \\
\hline
\end{tabular*}  
\label{T:5}
\footnote*{\footnotesize{EST: Parameter Estimate, SE: Standard Error, BIAS: Bias in Estimation, RMSE: Root Mean Squared Error, CI: Confidence Interval, CP: Coverage Probability (Nominal Level of 95\%)}}
\end{table}



\begin{table}[htbp]
\caption{EST, SE, bias, RMSE, 95$\%$ CI and CP for the destructive negative binomial cure model with $\phi=0.5$  based on $n=400$}
\begin{tabular*}{\textwidth}{c @{\extracolsep{\fill}}  c c c c c c c c c }
\hline
     ($p_{\min}, p_{\max}$) & $\psi$ & $\theta$ &     True Value &       EST &         SE &      BIAS &       RMSE &         95$\%$ CI &         CP  \\
\hline


        (0.2, 0.6) & 0.05 &   $\alpha$ &      1.099 &      1.061 &      0.246 &     -0.038 &      0.352 &       (0.580,      1.543) &      0.916 \\

       &&   $\beta_0$ &      -1.386 &     -1.432 &      0.248 &     -0.045 &      0.337 &     (-1.918,     -0.946) &      0.949 \\

       &&   $\beta_1$ &       0.115 &      0.116 &      0.071 &      0.013 &      0.101 &     (-0.022,      0.255) &      0.928 \\

       &&   $\gamma_0$ &       1.658 &      1.827 &      0.136 &      0.169 &      0.241 &      (1.561,      2.093) &      0.786 \\

       &&   $\gamma_1$ &      3.765 &      3.917 &      0.441 &      0.153 &       0.620 &      (3.052,      4.782) &      0.925 \\

       &&   $\gamma_2$ &     -0.005 &     -0.023 &       0.040 &     -0.018 &      0.057 &     (-0.102,      0.056) &      0.933 \\

   &&      $\gamma_3$ &      0.024 &     -0.131 &      0.262 &     -0.155 &      0.383 &     (-0.645,      0.382) &      0.897 \\

   &&      $\phi$ &        0.500 &      0.369 &     - &    - &      - &          - &          - \\

\hline  


        (0.3, 0.9) & 0.05 &   $\alpha$ &      1.099 &      1.012 &      0.212 &     -0.087 &       0.310 &      (0.597,      1.426) &      0.894 \\

       &&   $\beta_0$ &      -0.847 &     -0.905 &      0.255 &     -0.057 &      0.349 &     (-1.404,     -0.405) &      0.944 \\

       &&   $\beta_1$ &       0.159 &      0.194 &      0.107 &      0.017 &      0.159 &     (-0.016,      0.403) &      0.869 \\

       &&   $\gamma_0$ &       1.658 &      1.837 &      0.117 &       0.180 &      0.229 &      (1.607,      2.068) &      0.683 \\

       &&   $\gamma_1$ &      3.765 &      3.968 &      0.379 &      0.203 &      0.551 &      (3.225,       4.710) &      0.924 \\

       &&   $\gamma_2$ &     -0.005 &     -0.028 &      0.036 &     -0.023 &      0.053 &       (-0.100,      0.043) &      0.916 \\

       &&   $\gamma_3$ &      0.024 &     -0.215 &      0.231 &     -0.239 &      0.382 &     (-0.667,      0.238) &      0.806 \\

       &&   $\phi$ &        0.500 &      0.265 &      - &    - &      - &          - &          - \\

\hline


       (0.2, 0.6) & 0.15 &   $\alpha$ &      1.099 &      1.059 &       0.310 &      -0.040 &       0.430 &      (0.452,      1.667) &      0.931 \\

      &&   $\beta_0$ &      -1.387 &      -1.450 &      0.312 &     -0.063 &      0.417 &     (-2.062,     -0.838) &      0.959 \\

       &&   $\beta_1$ &       0.098 &      0.128 &      0.097 &      0.024 &      0.135 &     (-0.062,      0.318) &      0.931 \\

       &&   $\gamma_0$ &       1.658 &      1.816 &      0.162 &      0.158 &      0.264 &      (1.498,      2.133) &      0.855 \\

       &&   $\gamma_1$ &      3.765 &      3.873 &      0.615 &      0.109 &      0.833 &      (2.669,      5.078) &      0.959 \\

       &&   $\gamma_2$ &     -0.005 &     -0.024 &      0.057 &     -0.019 &      0.079 &     (-0.136,      0.088) &      0.935 \\

       &&   $\gamma_3$ &      0.024 &     -0.166 &      0.367 &      -0.190 &      0.531 &     (-0.886,      0.554) &       0.910 \\

       &&   $\phi$ &        0.500 &      0.381 &      - &     - &     - &          -  &          - \\

\hline

 (0.3, 0.9) & 0.15 & $\alpha$ &      1.099 &      1.048 &      0.271 &     -0.051 &      0.381 & (0.517, 1.579) &      0.928 \\

&& $\beta_0$ &     -0.847 &     -0.936 &      0.323 &     -0.088 &      0.442 & (-1.568, -0.304) &      0.941 \\

&& $\beta_1$ &      0.213 &      0.222 &      0.145 &      0.045 &      0.217 & (-0.062, 0.505) &      0.885 \\

&& $\gamma_0$ &      1.658 &      1.831 &      0.141 &      0.174 &      0.246 & (1.556, 2.107) &      0.774 \\

&& $\gamma_1$ &      3.765 &      3.936 &      0.526 &      0.171 &      0.724 & (2.905, 4.966) &      0.946 \\

&& $\gamma_2$ &     -0.005 &     -0.031 &       0.050 &     -0.025 &      0.073 & (-0.129, 0.068) &      0.906 \\

&& $\gamma_3$ &      0.024 &     -0.259 &      0.326 &     -0.283 &      0.502 & (-0.898, 0.381) &      0.871 \\

   && $\phi$ &        0.5 &      0.303 &          - &          - &          - &          - &          - \\

\hline

        (0.2, 0.6) & 0.25 &   $\alpha$ &      1.099 &      1.074 &        0.400 &     -0.025 &      0.537 &       (0.290,      1.857) &      0.946 \\

       &&   $\beta_0$ &      -1.387 &     -1.454 &      0.398 &     -0.067 &      0.529 &     (-2.234,     -0.674) &      0.957 \\

       &&   $\beta_1$ &       0.094 &      0.131 &      0.119 &      0.027 &      0.164 &     (-0.103,      0.365) &      0.945 \\

       &&   $\gamma_0$ &       1.658 &      1.826 &       0.190 &      0.169 &      0.297 &      (1.454,      2.199) &      0.882 \\

       &&   $\gamma_1$ &      3.765 &      3.856 &      0.841 &      0.092 &       1.140 &      (2.208,      5.504) &      0.922 \\

       &&   $\gamma_2$ &     -0.005 &     -0.027 &      0.077 &     -0.022 &      0.105 &     (-0.177,      0.123) &      0.933 \\

       &&   $\gamma_3$ &      0.024 &     -0.194 &      0.503 &     -0.218 &      0.695 &      (-1.180,      0.792) &      0.935 \\

       &&   $\phi$ &        0.500 &      0.399 &      -  &     - &      - &          - &          - \\

\hline


         (0.3, 0.9) & 0.25 &   $\alpha$ &      1.099 &      1.068 &      0.361 &     -0.031 &      0.492 &      (0.361 ,      1.775) &      0.936 \\

       &&   $\beta_0$ &      -0.847 &     -0.944 &      0.413 &     -0.096 &      0.553 &     (-1.754,     -0.133) &      0.944 \\

       &&   $\beta_1$ &       0.204 &       0.240 &      0.186 &      0.064 &      0.267 &     (-0.124,      0.604) &      0.895 \\

       &&   $\gamma_0$ &       1.658 &      1.824 &      0.164 &      0.166 &      0.266 &      (1.502,      2.146) &      0.856 \\

       &&   $\gamma_1$ &      3.765 &      3.935 &      0.733 &       0.170 &      0.976 &      (2.498,      5.372) &      0.951 \\

       &&   $\gamma_2$ &     -0.005 &     -0.029 &      0.065 &     -0.024 &      0.092 &     (-0.156,      0.097) &      0.912 \\

       &&   $\gamma_3$ &      0.024 &     -0.302 &      0.453 &     -0.326 &       0.660 &     (-1.191,     0.587) &      0.916 \\

       &&   $\phi$ &        0.500 &      0.327 &     -  &     - &     - &         - &          - \\

\hline  

\end{tabular*}  
\label{T:6}
\footnote*{\footnotesize{EST: Parameter Estimate, SE: Standard Error, BIAS: Bias in Estimation, RMSE: Root Mean Squared Error, CI: Confidence Interval, CP: Coverage Probability (Nominal Level of 95\%)} }
\end{table}

\hspace{5mm}As mentioned in Section \ref{sect4}, we estimate the parameters using the EM algorithm, except for $\phi$ which is estimated using the profile likelihood approach. The admissible ranges for $\phi$ are taken to be $\Phi=\{-2.00, -1.90, \ldots, 2.00 \}$ for the DEWP cure model with true $\phi=0.2$, whereas $\Phi=\{0.10, 0.15, \ldots, 2.00\}$ for the DNB cure model when true $\phi=0.5$  and $\Phi=\{3.0, 3.1, \ldots, 7.0\}$ when true $\phi=5.2$. Apart from $\phi$,  an initial parameter value is chosen uniformly from the interval ($0.85 \theta_r$, $1.15 \theta_r$) where $\theta_r$ denotes true value of the parameter. In Tables \ref{T:4} - \ref{T:6}, we display the results of our simulation study for some chosen scenarios.  The accuracy and robustness of the proposed method of estimation are assessed through average estimated value (EST), standard error (SE), bias (BIAS), root mean squared error (RMSE), 95 $\%$ Confidence Interval (CI) and coverage probability (CP). CPs are obtained by using the asymptotic normality of the ML estimators and a nominal level of 95$\%$ for the confidence interval thus constructed. The results are based on 500 replications of simulated data for each scenario and all calculations are performed in R-3.1.3.  \\

\hspace{5mm} From Tables \ref{T:4} - \ref{T:6}, we observe the estimates to be close to the true parameter values, and the bias are small implying high accuracy of the estimation method. The profile likelihood method seems to perform relatively well in terms of accuracy, when data are generated from  the DEWP ($\phi=0.2$) cure model. However, when the true model is  DNB, bias are found to be high for the estimates of $\phi$. It can be attributed to the fact that the likelihood function is very flat with respect to parameter $\phi$. Under-coverage for $\beta_0$ and $\gamma_0$ are observed for DEWP and DNB cure models, respectively. To explain this under-coverage, we consider one such setting in which data are generated from the DEWP model with $\phi=0.2$ having large sample size ($n=400$), $(p_{\min}, p_{\max})=(0.2, 0.6)$ and low censoring ($\psi=0.05$). We fit DEWP cure models to the data obtained from 100 replications and compare the effect of estimating $\phi$ versus fixed $\phi$  on the coverage probabilities of the other parameters. These results are presented in Table \ref{T:7} from which we observe that the coverage probability of $\beta_0$ reaches the nominal level of 95$\%$ when $\phi$ is not estimated. This immediately points towards the imprecision in estimating $\phi$ (likely due to the flatness of the likelihood surface) leading to the under-coverage of $\beta_0$. The SE and RMSE decrease with an increase in the sample size and decrease in the censoring percentages, as one would naturally expect. For brevity, tables corresponding to other scenarios are not provided here.\\

\begin{table}[htbp]
\caption{EST, SE, BIAS, RMSE, 95$\%$ CI and CP for destructive exponentially weighted Poisson cure model with $\phi=0.2$ with $(p_{\min}, p_{\max})=(0.2, 0.6)$ and $\psi=0.05$ based on $n=400$}
\begin{tabular*}{\textwidth}{c @{\extracolsep{\fill}}  c c c c c c c }
\hline
      $\theta$     & True Value &       EST &     SE &       BIAS &       RMSE  &   95$\%$CI &       CP \\
\hline
&\multicolumn{7}{c}{$\phi$ is estimated with $\hat \phi=0.597$}\\
\hline
 $\alpha$ &      1.099 &      1.064 &      0.187 &     -0.035 &      0.252 & (0.698, 1.430) &      0.929 \\

$\beta_0$ &     -1.386 &     -1.809 &      0.333 &     -0.422 &      1.452 & (-2.462, -1.156) &      0.291 \\

$\beta_1$ &      0.099 &      0.778 &      0.570 &      0.675 &      0.923 & (-0.339, 1.894) &      0.899 \\

$\gamma_0$ &      1.658 &      1.816 &      0.120 &      0.158 &      0.215 & (1.581, 2.050) &      0.758 \\

$\gamma_1$ &      3.765 &      3.953 &      0.391 &      0.188 &      0.555 & (3.187, 4.718) &      0.929 \\

$\gamma_2$ &     -0.005 &     -0.022 &      0.035 &     -0.017 &      0.049 & (-0.091, 0.047) &      0.919 \\

$\gamma_3$ &      0.024 &     -0.148 &      0.230 &     -0.172 &      0.357 & (-0.598, 0.302) &      0.848 \\

\hline
&\multicolumn{7}{c}{$\phi$ is not estimated}\\
\hline

$\alpha$ &      1.099 &      1.086 &      0.204 &     -0.013 &      0.264 & (0.685, 1.486) &      0.979 \\

$\beta_0$ &     -1.386 &     -1.348 &      0.208 &      0.038 &      0.276 & (-1.756, -0.941) &      0.979 \\

$\beta_1$ &      0.099 &      0.099 &      0.053 &     -0.004 &      0.069 & (-0.005, 0.202) &      0.989 \\

$\gamma_0$ &      1.658 &      1.815 &      0.120 &      0.157 &      0.214 & (1.581, 2.049) &      0.778 \\

$\gamma_1$ &      3.765 &      3.944 &      0.390 &      0.180 &      0.553 & (3.179, 4.709) &      0.959 \\

$\gamma_2$ &     -0.005 &     -0.022 &      0.036 &     -0.016 &      0.050 & (-0.092, 0.049) &      0.939 \\

$\gamma_3$ &      0.024 &     -0.154 &      0.230 &     -0.178 &      0.362 & (-0.606, 0.297) &      0.870 \\

\hline
\end{tabular*}  
\label{T:7}
\footnote*{\footnotesize{EST: Parameter Estimate, SE: Standard Error, BIAS: Bias in Estimation, RMSE: Root Mean Squared Error, CI: Confidence Interval, CP: Coverage Probability (Nominal Level of 95\%)} }
\end{table}

\section{Model discrimination}\label{sect7}
\hspace{5mm}  To assess the impact of model mis-specification on estimates of the cure rate, a model discrimination is performed here based on Akaike Information Criterion (AIC) and Bayesian Information Criterion (BIC). The idea is to observe the frequency with which models other than the true model get selected or rejected  by our fitting method of these models.  We generate 1000 samples from five true models each, {\it viz.}, DEWP ($\phi=-0.5$), DEWP ($\phi=0.2$), DLBP, DNB ($\phi=0.5$) and DNB ($\phi=0.75$) with $(p_{\min}, p_{\max})=(0.3, 0.9)$, $\eta=3$ for $Z=1$ and $\lambda=0.15$ (i.e., medium censoring). The lifetime parameters are set as $\bm \gamma=(\gamma_0, \gamma_1, \gamma_2, \gamma_3)'=(1.657, 3.764, -0.005, 0.023)'$ (see Section \ref{sect6}). Under these specifications, samples are generated of size $n=400$.   \\

\hspace{5mm}  We fit three candidate models, i.e.,  DEWP, DLBP and DNB cure models, to these samples with the proposed method of estimation.  The model with the least AIC or BIC value gets selected, where $$ \text{AIC}= -2 \hat l + 2 \tilde k, \hspace{0.4cm}  \text{BIC} = -2 \hat l + \tilde k \log(n), $$ with $\hat l$ being the maximized log-likelihood value corresponding to the model and $\tilde k$ being the  number of estimated parameters. The selection rates based on AIC, BIC and $\hat l$ are all presented in Table \ref{T:8}. \\

\begin{table}[htbp]
\caption{Selection rates based on AIC, BIC and maximized log-likelihood ($\hat l$) value for $n=400$}
\begin{tabular*}{\textwidth}{c @{\extracolsep{\fill}}ccc}
           &          \multicolumn{ 3}{c}{Fitted Models}  \\
\cline{2-4}
           True Models   & DEWP     & DLB & DNB   \\
\hline
\multicolumn{ 1}{c}{DEWP ($\phi=-0.5$)} & $\hat \phi=-0.275$ &            &  $\hat \phi=0.378$ \\
\cline{1-4}
\multicolumn{ 1}{c}{AIC} &      0.179 &      0.768 &      0.053 \\

\multicolumn{ 1}{c}{BIC}  &      0.037 &      0.944 &      0.019 \\

\multicolumn{ 1}{c}{log-lik} &      0.630 &      0.152 &      0.218 \\
\hline
\multicolumn{ 1}{c}{DEWP ($\phi=0.2$)} &  $\hat \phi=0.222$ &            &  $\hat \phi=0.186$ \\
\cline{1-4}
\multicolumn{ 1}{c}{AIC}  &      0.125 &      0.843 &      0.032 \\

\multicolumn{ 1}{c}{BIC}  &      0.063 &      0.919 &      0.018 \\

\multicolumn{ 1}{c}{log-lik}  &      0.597 &      0.360 &      0.043 \\
\hline

\multicolumn{ 1}{c}{DLB} & $\hat \phi=-0.077$ &            &  $\hat \phi=0.347$ \\
\hline
\multicolumn{ 1}{c}{AIC}  &      0.073 &      0.919 &      0.008 \\

\multicolumn{ 1}{c}{BIC}  &      0.016 &      0.983 &      0.001 \\

\multicolumn{ 1}{c}{log-lik}  &      0.427 &      0.559 &      0.014 \\

\hline
\multicolumn{ 1}{c}{DNB ($\phi=0.5$)}  &  $\hat \phi=0.311$ &            &  $\hat \phi=0.336$ \\
\hline
\multicolumn{ 1}{c}{AIC} &      0.163 &      0.762 &      0.075 \\

\multicolumn{ 1}{c}{BIC} &      0.003 &      0.969 &      0.028 \\

\multicolumn{ 1}{c}{log-lik}  &      0.556 &      0.262 &      0.182 \\
\hline
\multicolumn{ 1}{c}{DNB ($\phi=0.75$)}  &  $\hat \phi=0.545$ &            &  $\hat \phi=0.346$ \\
\hline
\multicolumn{ 1}{c}{AIC}  &      0.174 &      0.737 &      0.089 \\

\multicolumn{ 1}{c}{BIC}  &      0.040 &      0.927 &      0.033 \\

\multicolumn{ 1}{c}{log-lik}  &      0.599 &      0.242 &      0.159 \\
\hline
\end{tabular*}  
\label{T:8}
\footnote*{\footnotesize{AIC: Akaike Information Criterion, BIC: Bayesian Information Criterion, log-lik: Maximized log-likelihood value }}
\end{table}

\hspace{5mm}  The selection of true models based on both AIC and BIC values are found to be quite low when the data are generated from DEWP and DNB cure models.  It is observed that the values of the log-likelihood function for all the fitted cure models are quite similar. For  this reason, the models with one extra parameter (i.e. DEWP and DNB), in terms of AIC and BIC values, do get penalized more when compared to the DLB model. Consequently, when the log-likelihood value ($\hat l$) is used as the selection criteria, the results reveal more selections for the true models. Table \ref{T:9} shows that when $\phi$ is not estimated, true models are more likely to get selected.   \\

\begin{table}[htbp]
\caption{Comparison of model selection rates based on AIC for $n=400$}
\begin{tabular*}{\textwidth}{l @{\extracolsep{\fill}}cc}
\multicolumn{ 1}{c}{ } &                  \multicolumn{ 2}{c}{True Model} \\
\cline{2-3}
\multicolumn{ 1}{c}{Fitted Model} & \multicolumn{ 1}{c}{DEWP ($\phi=-0.5$)} & \multicolumn{ 1}{c}{DNB ($\phi=0.5$)} \\
\hline
True Model  &          0.540 &       0.330  \\

      DEWP &            0.060 &        0.100  \\

      DLBP &              0.390 &       0.530  \\

       DNB &              0.010 &       0.040  \\
\hline
\end{tabular*}  
\label{T:9}
\footnote*{\footnotesize{AIC: Akaike Information Criterion}}
\end{table}

\hspace{5mm}  To emphasize the importance of model discrimination, we examine the bias and MSE involved in the estimation of cure rates of patients under model mis-specification. For each model, we compute the total relative bias (TRB) as

$$ \text{TRB} = \sum_{i=1}^n \frac{|\hat \pi_{0,i}-\pi_{0,i}|}{\pi_{0,i}},$$ 
where $\pi_{0,i}$ and $\hat \pi_{0,i}$ denote true and estimated cure rates for an individual $i$, $i=1, \dots, n$. Similarly, we define total mean squared error (TMSE) for a model as
$$ \text{TMSE}=\frac{1}{n-1}\sum_{i=1}^n (\hat \pi_{0,i}-\pi_{0,i})^2.$$
For two candidate models M1 and M2, total relative efficiency (TRE) of M2 with respect to M1 is defined as $\text{TRE}=\frac{\text{TMSE(M2)}}{\text{TMSE(M1)}}$, where $\text{TMSE(M1)}$ and $\text{TMSE(M2)}$ denote TMSE  values based on  M1 and M2, respectively.  With these measures defined, we compare the three candidate models. Table \ref{T:10} presents TRB (in $\%$), TMSE and TRE for the candidate models for $n=400$ when the data are generated from one of the five true models as described earlier in this section. \\

\begin{table}[h!]
\begin{center}
\caption{TRB $(\%)$ (TMSE, $\hat \phi$, TRE) in estimation of cured proportion for all candidate models for $n=400$}
\resizebox{16.5cm}{!}{
\begin{tabular}{lcccc}
\multicolumn{ 1}{c}{} &                               \multicolumn{ 4}{c}{Fitted Model} \\
\cline{2-5}
\multicolumn{ 1}{c}{True Model} & True Model & DEWP & DLBP &  DNB \\
\hline
DEWP($\phi=-0.5$) & 35.300 (0.003, -, 1.000) & 37.015 (0.004, -0.199, 0.962) & 37.730 (0.004, -, 1.383)  & 34.957 (0.003, 0.461, 1.045)\\
DEWP($\phi=0.2$) & 62.365 (0.003, - , 1.000) & 66.532 (0.004, 0.239, 1.004) & 61.101 (0.003, - , 1.087) & 67.786 (0.003, 0.198, 1.094) \\
DLBP & 86.617 (0.003, - , 1.000) &107.147 (0.004, 0.708, 0.964) & 86.617 (0.003, - , 1.000) & 193.413 (0.006, 0.117, 0.455)\\
DNB ($\phi=0.5$) &  41.663 (0.004, - , 1.000)  & 42.593 (0.004, -0.079, 1.006) & 42.992 (0.004, - , 1.052) & 40.039 (0.004, 0.396, 1.123) \\
DNB ($\phi=0.75$) & 37.100 (0.003, - , 1.000) & 39.126 (0.004, -0.259, 0.972) & 40.846 (0.004, - , 1.047) &37.247 (0.003, 0.379, 1.030)\\
\hline
\end{tabular}  
}
\raggedright \footnote*{\raggedleft {\footnotesize{TRB: Total Relative Bias, TMSE: Total Mean Squared Error, TRE: Total Relative Efficiency}}}
\label{T:10}
\end{center}
\end{table}

\hspace{5mm}  The model M1 gets chosen always to be the true model. From Table \ref{T:10}, it can be seen that in cases where data are from the DLBP cure model, model mis-specification may lead to large bias and MSE, and consequently, higher TRB and lower TRE are observed on fitting candidate models when the true model is DLBP.  For the other true models, TRB values are relatively close to each other which indicate not much precision is lost under model mis-specification. DNB cure models provide lesser TRB and higher TRE in most of the scenarios considered. Table \ref{T:11} shows TRB and TRE values when using AIC and $\hat l$ as the model selection
criteria. The results suggest that  allowing AIC or $\hat l$ to select a working model out of a set of candidate models may lead to lesser relative bias. TRE values are greater than one in most cases, which means that estimating the cured proportion
by fitting the working model as selected by AIC or $\hat l$ results in higher efficiency.

\begin{table}[htbp]
\caption{TRB ($\%$) and TRE when AIC and $\hat l$ are used for model selection for $n=400$}
\begin{tabular*}{\textwidth}{l @{\extracolsep{\fill}}cccc}
             & \multicolumn{ 2}{c}{AIC} & \multicolumn{ 2}{c}{$\hat l$}  \\
\cline{2-5}
              True Model  & TRB ($\%$) &        TRE & TRB ($\%$) &        TRE \\
\hline
DEWP ($\phi=-0.5$)  &     36.347 &      1.148 &     36.659 &      1.085 \\

DEWP ($\phi=0.2$) &     62.321 &       1.040 &     63.832 &      1.032 \\

       DLB &         88.259 &      0.997 &     94.829 &      0.986 \\

DNB ($\phi=0.5$) &     42.461 &       1.030 &     42.408 &      1.027 \\

DNB ($\phi=0.75$) &     39.347 &      1.023 &     39.104 &      0.998 \\
\hline
\end{tabular*}
\footnote*{{\footnotesize{TRB: Total Relative Bias,  TRE: Total Relative Efficiency}}}  
\label{T:11}
\end{table}

\section{Concluding Remarks} \label{sect8}

\hspace{5mm} In this work, destructive cure models are studied under competing risks scenario wherein the initial competing causes undergo a destructive mechanism. The models are developed and examined assuming that the hazard functions corresponding to the susceptible individuals follow proportional hazards with Weibull baseline hazard function. The model generalizes the previous works of  \cite{ pal2016destructive, pal2016likelihood} on destructive cure model by assuming non i.i.d. lifetimes for the susceptible individuals. This is accomplished by linking covariates to the lifetimes through proportional hazards. The parameter estimates are found to be accurate with low RMSE. A relatively large bias is observed while estimating $\phi$, especially when data are from DNB ($\phi=0.75$) cure model. The estimates are observed to be more precise under scenarios characterized by low censoring, high proportion of undamaged competing causes and large sample size. A model discrimination is also carried out using information-based criteria.  A real life example on cutaneous melanoma is considered for the purpose of illustrating the models and the method of fit developed here. Destructive negative binomial cure model with $\hat \phi=5.2$ provided the best fit to these melanoma data. The assumption of  i.i.d. lifetimes among the susceptible subjects got rejected at $10\%$ level of significance. It will be of interest to study destructive cure models when covariates are prone to measurement errors. We hope to consider this problem as our future research.

\bibliographystyle{apacite}
\bibliography{dcm}

\vspace{1cm}
\appendix
{\LARGE \bf Appendix}
\section{Q-function}\label{A3}
\setcounter{equation}{0}
\numberwithin{equation}{section}
\hspace{5mm}For all $i \in \{1, 2, \dots, n\}$ and following Eqs. (\ref{eqx21}) - (\ref{eqx24}), we define

$$\eta_i=e^{\bm \alpha' \bm z_i},\text{ } p_i=\frac{e^{\bm \beta' \bm x_i}}{1+e^{\bm \beta' \bm x_i}},$$
$$\lambda_i=\lambda(t_i; \bm x_i, \bm z_i, \bm \gamma), \text{ }  S_i=S(t_i; \bm x_i, \bm z_i, \bm \gamma), \text{ }  F_i=F(t_i; \bm x_i, \bm z_i, \bm \gamma) \text{ and } f_i=f(t_i; \bm x_i, \bm z_i, \bm \gamma).$$

\subsection{Destructive exponentially weighted Poisson cure model}
\begin{flalign*}
Q\left(\bm\theta^*, \bm \xi^{(a)}\right)=\sum_{\Delta_1} \log M_i - \sum_{i=1}^n M_i+\sum_{\Delta_1}M_iS_i +\sum_{\Delta_1} \log f_i+\sum_{\Delta_0}\xi^{(a)}_{i}\log\left(e^{M_iS_i}-1\right),
\end{flalign*}
where
$$ \xi^{(a)}_{i}=1-e^{-\eta_i e^{\phi} p_i S_i}\bigg \rvert_{\bm \theta^*=\hat{\bm \theta}^{*(a)}} \text{ and } M_i =  \eta_ie^{\phi}p_i.$$ 

\subsection{Destructive length-biased Poisson cure model}
\begin{flalign*}
 Q\left(\bm\theta^*, \bm \xi^{(a)}\right)=\sum_{\Delta_1} \log {\eta_i}+\sum_{\Delta_1}\log{p_i}+\sum_{\Delta_1}\log{f_i}-\sum_{\Delta_1}{A_i}
+\sum_{\Delta_1}{B_i}\\-\sum_{\Delta_0}\eta_i p_i + \sum_{\Delta_0}\log(1-p_i)+\sum_{\Delta_0} \xi_i^{(a)} \log(C_iD_i-1),
\end{flalign*}
where
$$ \xi^{(a)}_{i}=1-e^{-\eta_i  p_i S_i}\left(\frac{1-p_i}{1-p_iF_i}\right)\bigg \rvert_{\bm \theta^*=\hat{\bm \theta}^{*(a)}},$$
$$A_i=\eta_i p_i F_i, \text{ }  B_i=\log\left(1-p_iF_i-\frac{p_if_i}{\eta_i}\right), \text{ }  C_i=e^{\eta_i p_i (1-F_i)} \text{ and } D_i=\frac{1-p_iF_i}{1-p_i}.$$  
\subsection{Destructive negative binomial cure model}
\begin{flalign*}
 Q\left(\bm\theta^*, \bm \xi^{(a)}\right)=\sum_{\Delta_1} \log \eta_i p_i -\left(\frac{1}{\phi}+1\right) \sum_{\Delta_1}\log(1+E_iF_i) +\sum_{\Delta_1}\log f_i &\\-\frac{1}{\phi}\sum_{\Delta_0} \log (1+E_i)+\sum_{\Delta_0}\xi^{(a)}_{i}\log \left(G_i^{-1/\phi}-1\right),
\end{flalign*}
where
$$ \xi^{(a)}_{i}= 1- G_i \bigg \rvert_{\bm \theta^*= \hat{\bm \theta}^{*(a)}}, \text{ }  E_i = \phi \eta_i p_i 
\text{ and }
G_i=\frac{1+\phi \eta_i p_iF_i}{1+\phi \eta_i p_i}.$$ 

\section{First- and second-order derivatives}\label {B3}

\subsection{Cumulative distribution function}
$$F'_{i,0}=\frac{\partial F_i}{\partial \gamma_0}=-S_i\log S_i\log\left(\frac{t_i}{\gamma_1}\right), \quad F'_{i,1}=\frac{\partial F_i}{\gamma_1}=S_i\log S_i\log\left(\frac{\gamma_0}{\gamma_1}\right),$$
$$F'_{i,2l}=\frac{\partial F_i}{\partial \gamma_{2l}}=-x_{il}S_i\log S_i,  \quad  F'_{i,3m}=\frac{\partial F_i}{\gamma_{3m}}=-z_{im}S_i\log S_i ,$$
$$F''_{i,00}=\frac{\partial^2 F_i}{\partial \gamma_0^2}=-\left[\log\left(\frac{t_i}{\gamma_0}\right)\right]^2 S_i \log S_i(1+\log S_i), \quad F''_{i,01}=\frac{\partial^2 F_i}{\partial \gamma_0 \partial \gamma_1}=\frac{S_i\log S_i}{\gamma_1}\left[1+\gamma_0\log\left(\frac{t_i}{\gamma_1}\right)(1+\log S_i)\right],$$
$$F''_{i,11}=\frac{\partial^2 F_i}{\partial \gamma_1^2 }=-\frac{\gamma_0}{\gamma_1^2} S_i\log S_i\left[1+\gamma_0\log\left(\frac{t_i}{\gamma_1}\right)\right],  \quad F''_{i,0(2l)}=\frac{\partial^2 F_i}{\partial \gamma_0 \partial \gamma_{2l} }=-x_{il}\log\left(\frac{t_i}{\gamma_1}\right) S_i\log S_i(1+\log S_i),$$
$$F''_{i,0(3m)}=\frac{\partial^2 F_i}{\partial \gamma_0 \partial \gamma_{3m} }=-z_{im}\log\left(\frac{t_i}{\gamma_1}\right) S_i\log S_i(1+\log S_i), \quad F''_{i,1(2l)}=\frac{\partial^2 F_i}{\partial \gamma_1 \partial \gamma_{2l} }=x_{il}\left(\frac{\gamma_0}{\gamma_1}\right) S_i\log S_i(1+\log S_i),$$
$$F''_{i,1(3m)}=\frac{\partial^2 F_i}{\partial \gamma_1 \partial \gamma_{3m} }=z_{im}\left(\frac{\gamma_0}{\gamma_1}\right) S_i\log S_i(1+\log S_i), \quad F''_{i,(2l)(2l')}=\frac{\partial^2 F_i}{\partial \gamma_{2l} \partial \gamma_{2l'} }=-x_{il}x_{il'} S_i\log S_i(1+\log S_i),$$
$$F''_{i,(2l)(3m)}=\frac{\partial^2 F_i}{\partial \gamma_{2l} \partial \gamma_{3m} }=-x_{il}z_{im} S_i\log S_i(1+\log S_i),$$ and  
$$F''_{i,(3m)(3m')}=\frac{\partial^2 F_i}{\partial \gamma_{3m} \partial \gamma_{3m'} }=-z_{im}z_{im'} S_i\log S_i(1+\log S_i),$$
for
$i=1, \dots, n$; $j, j'= 1, \dots, q_1$; $k, k'= 0, 1, \dots, q_2$; $r, r'= 0,1, 20, 21, \dots, 2q_2, 31, 32, \dots, 3q_1$; $l, l'= 0, 1, \dots, q_2$; $m, m'=  1, \dots, q_1$ and $x_{i0}\equiv 1.$

\subsection{Survival function}
$$S'_{i,0}=-F'_{i,0}, \quad S'_{i,1}=-F'_{i,1}, \quad  S'_{i,2l}=-F'_{i,2l}, \quad  S'_{i,3m}=-F'_{i,3m}, \quad  S''_{i,00}=-F''_{i,00}, \quad  S''_{i,01}=-F''_{i,01},$$ 
$$S''_{i,0(2l)}=-F''_{i,0(2l)}, \quad  S''_{i,0(3m)}=-F''_{i,0(3m)}, \quad  S''_{i,11}=-F''_{i,11}, \quad  S''_{i,1(2l)}=-F''_{i,1(2l)}, \quad  S''_{i, 1(3m)}=-F''_{i, 1(3m)},$$
$$S''_{i,(2l)(2l')}=-F''_{i,(2l)(2l')}, \quad  S''_{i,(2l)(3m)}=-F''_{i,(2l)(3m)}, \text{ and }  S''_{i, (3m)(3m')}=-F''_{i, (3m)(3m')}.$$
for
$i=1, \dots, n$; $j, j'= 1, \dots, q_1$; $k, k'= 0, 1, \dots, q_2$; $r, r'= 0,1, 20, 21, \dots, 2q_2, 31, 32, \dots, 3q_1$; $l, l'= 0, 1, \dots, q_2$; $m, m'=  1, \dots, q_1$ and $x_{i0}\equiv 1.$

\subsection{Probability density function}
$$f'_{i,0}=\frac{\partial f_i}{\partial \gamma_0}=\left\{-F'_{i,0}+S_i\left[\frac{1}{\gamma_0}+ \log\left(\frac{t_i}{\gamma_1}\right)\right]\right\}\lambda_i, \quad  f'_{i,1}=\frac{\partial f_i}{\gamma_1}=-\left\{F'_{i,1}+S_i\left(\frac{\gamma_0}{\gamma_1}\right)\right\}\lambda_i,$$
$$f'_{i,2l}=\frac{\partial f_i}{\partial \gamma_{2l}}=\left\{-F'_{i,2l}+S_i x_{il}\right\}\lambda_i, \quad f'_{i,3m}=\frac{\partial f_i}{\partial \gamma_{3m}}=\left\{-F'_{i,3m}+S_i z_{im}\right\}\lambda_i ,$$
$$f''_{i,00}=\frac{\partial^2 f_i}{\partial \gamma_0^2}=\left\{- F''_{i,00}+ S_i \log \left(\frac{t_i}{\gamma_1}\right) \left[\frac{2}{\gamma_0}+\log\left( \frac{t_i}{\gamma_1}\right)\right]-2\left[\frac{1}{\gamma_0}+ \log\left(\frac{t_i}{\gamma_1}\right)\right]F'_{i,0}\right\}\lambda_i,$$
$$f''_{i,01}=\frac{\partial^2 f_i}{\partial \gamma_0 \partial \gamma_1}=\left\{- F''_{i,01}- S_i  \frac{\gamma_0}{\gamma_1} \left[\frac{2}{\gamma_0}+\log\left( \frac{t_i}{\gamma_1}\right)\right]-\left[\frac{1}{\gamma_0}+ \log\left(\frac{t_i}{\gamma_1}\right)\right] F'_{i,1}+\frac{\gamma_0}{\gamma_1} F'_{i,0}\right\}\lambda_i,$$
$$f''_{i,11}=\frac{\partial^2 f_i}{\partial \gamma_1^2}=\left\{- F''_{i,11}+ S_i \frac{\gamma_0(\gamma_0+1)}{\gamma^2_1}+2 \frac{\gamma_0}{\gamma_1}F'_{i,1}\right\}\lambda_i,$$
$$f''_{i,0(2l)}=\frac{\partial^2 f_i}{\partial \gamma_0 \partial \gamma_{2l} }=\left\{- F''_{i,0(2l)}+ (S_i x_{il}-F'_{i,2l})  \left[\frac{1}{\gamma_0}+ \log\left(\frac{t_i}{\gamma_1}\right)\right] -x_{il} F'_{i,0}\right\}\lambda_i,$$
$$f''_{i,0(3m)}=\frac{\partial^2 f_i}{\partial \gamma_0 \partial \gamma_{3m} }=\left\{- F''_{i,0(3m)}+ (S_i z_{im}-F'_{i,3m})  \left[\frac{1}{\gamma_0}+ \log\left(\frac{t_i}{\gamma_1}\right)\right] -z_{im} F'_{i,0}\right\}\lambda_i,$$
$$f''_{i,1(2l)}=\frac{\partial^2 f_i}{\partial \gamma_1 \partial \gamma_{2l} }=\left\{- F''_{i,1(2l)}- (S_i x_{il}-F'_{i,2l})  \frac{\gamma_0}{\gamma_1} -x_{il} F'_{i,1}\right\}\lambda_i,$$
$$f''_{i,1(3m)}=\frac{\partial^2 f_i}{\partial \gamma_1 \partial \gamma_{3m} }=\left\{- F''_{i,1(3m)}- (S_i z_{im}-F'_{i,3m})  \frac{\gamma_0}{\gamma_1} -z_{im} F'_{i,1}\right\}\lambda_i,$$
$$f''_{i,(2l)(2l')}=\frac{\partial^2 f_i}{\partial \gamma_{2l} \partial \gamma_{2l'} }=\left\{- F''_{i,(2l)(2l')}+ S_i x_{il}x_{il'}-F'_{i,2l}x_{il'}-x_{il} F'_{i,2l'}\right\}\lambda_i,$$
$$f''_{i,(2l)(3m)}=\frac{\partial^2 f_i}{\partial \gamma_{2l} \partial \gamma_{3m} }=\left\{- F''_{i,(2l)(3m)}+ S_i x_{il}z_{im}-F'_{i,2l}z_{im}-x_{il} F'_{i,3m}\right\}\lambda_i,$$
$$f''_{i,(3m)(3m')}=\frac{\partial^2 f_i}{\partial \gamma_{3m} \partial \gamma_{3m'} }=\left\{- F''_{i,(3m)(3m')}+ S_i z_{im}z_{im'}-F'_{i,3m}z_{im'}-z_{im} F'_{i,3m'}\right\}\lambda_i,$$
and
$$\frac{\partial \log f_i}{\partial \gamma_r}=\frac{f'_{i,r}}{f_i}, \frac{\partial^2 \log f_i}{\partial \gamma_r \partial \gamma_{r'}}=\frac{f_i f''_{i,rr'}-f'_{i,r}f'_{i,r'}}{f^2_i},$$
for
$i=1, \dots, n$; $j, j'= 1, \dots, q_1$; $k, k'= 0, 1, \dots, q_2$; $r, r'= 0,1, 20, 21, \dots, 2q_2, 31, 32, \dots, 3q_1$; $l, l'= 0, 1, \dots, q_2$; $m, m'=  1, \dots, q_1$ and $x_{i0}\equiv 1.$

\section{First- and second-order derivatives of  Q-function}\label{C3}
\hspace{5mm} For $i \in \{1, 2, \dots, n\}$, we define $D^*_i=\frac{e^{M_iS_i}}{e^{M_iS_i}-1}.$  

\subsection{Destructive exponentially weighted Poisson cure model}
\begin{flalign*}
\frac{\partial Q\left(\bm\theta^*, \bm \xi^{(a)}\right)}{\partial \alpha_j}=\sum_{\Delta_1}z_{ij}-\sum_{i=1}^nz_{ij}M_i+\sum_{\Delta_1}z_{ij}M_iS_i+\sum_{\Delta_0}\xi_i^{(a)}z_{ij}D^*_iM_iS_i,
\end{flalign*}
\begin{flalign*}
\frac{\partial Q\left(\bm\theta^*, \bm \xi^{(a)}\right)}{\partial \beta_k}=\sum_{\Delta_1}x_{ik}(1-p_i)-\sum_{i=1}^nx_{ik}M_i(1-p_i)+\sum_{\Delta_1}x_{ik}M_iS_i(1-p_i)+\sum_{\Delta_0}\xi_i^{(a)}x_{ik}D^*_iM_iS_i(1-p_i),
\end{flalign*}
\begin{flalign*}
\frac{\partial Q\left(\bm\theta^*, \bm \xi^{(a)}\right)}{\partial \gamma_0}=\sum_{\Delta_1}M_iS'_{i,0}+\sum_{\Delta_1}\left[\frac{1}{\gamma_0}+\log\left(\frac{t_i}{\gamma_1}\right)  + \frac{S'_{i,0}}{S_i}  \right]+\sum_{\Delta_0}\xi_i^{(a)}D^*_iM_iS'_{i,0},
\end{flalign*}
\begin{flalign*}
\frac{\partial Q\left(\bm\theta^*, \bm \xi^{(a)}\right)}{\partial \gamma_1}=\sum_{\Delta_1}M_iS'_{i,1}+\sum_{\Delta_1}\left[-\frac{\gamma_0}{\gamma_1}+ \frac{S'_{i,1}}{S_i}  \right]+\sum_{\Delta_0}\xi_i^{(a)}D^*_iM_iS'_{i,1},
\end{flalign*}
\begin{flalign*}
\frac{\partial Q\left(\bm\theta^*, \bm \xi^{(a)}\right)}{\partial \gamma_{2l}}=\sum_{\Delta_1}M_iS'_{i,2l}+\sum_{\Delta_1}\left[x_{il}+ \frac{S'_{i,2l}}{S_i}  \right]+\sum_{\Delta_0}\xi_i^{(a)}D^*_iM_iS'_{i,2l},
\end{flalign*}
\begin{flalign*}
\frac{\partial Q\left(\bm\theta^*, \bm \xi^{(a)}\right)}{\partial \gamma_{3m}}=\sum_{\Delta_1}M_iS'_{i,3m}+\sum_{\Delta_1}\left[z_{im} + \frac{S'_{i,3m}}{S_i}  \right]+\sum_{\Delta_0}\xi_i^{(a)}D^*_iM_iS'_{i,3m},
\end{flalign*}
\begin{flalign*}
\frac{\partial^2 Q\left(\bm\theta^*, \bm \xi^{(a)}\right)}{\partial \alpha_j \partial \alpha_{j'}}=-\sum_{i=1}^nz_{ij}z_{ij'}M_i+\sum_{\Delta_1}z_{ij}z_{ij'}M_iS_i+\sum_{\Delta_0}\xi_i^{(a)}z_{ij}z_{ij'}D^*_iM_iS_i\left[1-\frac{M_iS_i}{e^{M_iS_i}-1}\right],
\end{flalign*}
\begin{flalign*}
\frac{\partial^2 Q\left(\bm\theta^*, \bm \xi^{(a)}\right)}{\partial \alpha_j \partial \beta_{k}}=-\sum_{i=1}^nx_{ij}z_{ik}M_i(1-p_i)+\sum_{\Delta_1}x_{ik}z_{ij}M_iS_i(1-p_i)+\sum_{\Delta_0}\xi_i^{(a)}x_{ik}z_{ij}D^*_iM_iS_i(1-p_i)\left[1-\frac{M_iS_i}{e^{M_iS_i}-1}\right],
\end{flalign*}
\begin{flalign*}
\frac{\partial^2 Q\left(\bm\theta^*, \bm \xi^{(a)}\right)}{\partial \beta_k \partial \beta_{k'}}=-\sum_{i=1}^nx_{ik}x_{ik'}M_i(1-p_i)(1-2p_i)+\sum_{\Delta_1}x_{ik}x_{ik'}M_iS_i(1-p_i)(1-2p_i)\\+\sum_{\Delta_0}\xi_i^{(a)}x_{ik}x_{ik'}D^*_iM_iS_i(1-p_i)(1-2p_i)\left[1-\frac{M_iS_i(1-p_i)}{(1-2p_i)(e^{M_iS_i}-1)}\right],
\end{flalign*}
\begin{flalign*}
\frac{\partial^2 Q\left(\bm\theta^*, \bm \xi^{(a)}\right)}{\partial \alpha_j \partial \gamma_{0}}=\sum_{\Delta_1}z_{ij}M_iS'_{i,0}+\sum_{\Delta_0}\xi_i^{(a)}z_{ij}D^*_iM_iS'_{i,0}\left[1-\frac{M_iS_i}{e^{M_iS_i}-1}\right],
\end{flalign*}
\begin{flalign*}
\frac{\partial^2 Q\left(\bm\theta^*, \bm \xi^{(a)}\right)}{\partial \alpha_j \partial \gamma_{1}}=\sum_{\Delta_1}z_{ij}M_iS'_{i,1}+\sum_{\Delta_0}\xi_i^{(a)}z_{ij}D^*_iM_iS'_{i,1}\left[1-\frac{M_iS_i}{e^{M_iS_i}-1}\right],
\end{flalign*}
\begin{flalign*}
\frac{\partial^2 Q\left(\bm\theta^*, \bm \xi^{(a)}\right)}{\partial \alpha_j \partial \gamma_{2l}}=\sum_{\Delta_1}z_{ij}M_iS'_{i,2l}+\sum_{\Delta_0}\xi_i^{(a)}z_{ij}D^*_iM_iS'_{i,2l}\left[1-\frac{M_iS_i}{e^{M_iS_i}-1}\right],
\end{flalign*}
\begin{flalign*}
\frac{\partial^2 Q\left(\bm\theta^*, \bm \xi^{(a)}\right)}{\partial \alpha_j \partial \gamma_{3m}}=\sum_{\Delta_1}z_{ij}M_iS'_{i,3m}+\sum_{\Delta_0}\xi_i^{(a)}z_{ij}D^*_iM_iS'_{i,3m}\left[1-\frac{M_iS_i}{e^{M_iS_i}-1}\right],
\end{flalign*}
\begin{flalign*}
\frac{\partial^2 Q\left(\bm\theta^*, \bm \xi^{(a)}\right)}{\partial \beta_k \partial \gamma_{0}}=\sum_{\Delta_1}x_{ik}(1-p_i)M_iS'_{i,0}+\sum_{\Delta_0}\xi_i^{(a)}x_{ik}(1-p_i)D^*_iM_iS'_{i,0}\left[1-\frac{M_iS_i}{e^{M_iS_i}-1}\right],
\end{flalign*}
\begin{flalign*}
\frac{\partial^2 Q\left(\bm\theta^*, \bm \xi^{(a)}\right)}{\partial \beta_k \partial \gamma_{1}}=\sum_{\Delta_1}x_{ik}(1-p_i)M_iS'_{i,1}+\sum_{\Delta_0}\xi_i^{(a)}x_{ik}(1-p_i)D^*_iM_iS'_{i,1}\left[1-\frac{M_iS_i}{e^{M_iS_i}-1}\right],
\end{flalign*}
\begin{flalign*}
\frac{\partial^2 Q\left(\bm\theta^*, \bm \xi^{(a)}\right)}{\partial \beta_k \partial \gamma_{2l}}=\sum_{\Delta_1}x_{ik}(1-p_i)M_iS'_{i,2l}+\sum_{\Delta_0}\xi_i^{(a)}x_{ik}(1-p_i)D^*_iM_iS'_{i,2l}\left[1-\frac{M_iS_i}{e^{M_iS_i}-1}\right],
\end{flalign*}
\begin{flalign*}
\frac{\partial^2 Q\left(\bm\theta^*, \bm \xi^{(a)}\right)}{\partial \beta_k \partial \gamma_{3m}}=\sum_{\Delta_1}x_{ik}(1-p_i)M_iS'_{i,3m}+\sum_{\Delta_0}\xi_i^{(a)}x_{ik}(1-p_i)D^*_iM_iS'_{i,3m} \left[1-\frac{M_iS_i}{e^{M_iS_i}-1}\right],
\end{flalign*}

\begin{flalign*}
\frac{\partial^2 Q\left(\bm\theta^*, \bm \xi^{(a)}\right)}{\partial \gamma^2_0 }=\sum_{\Delta_1}M_iS''_{i,00}+\sum_{\Delta_1}\left[-\frac{1}{\gamma^2_0}+\frac{S_iS''_{i,00}-(S'_{i,0})^2}{S_i^2}\right]+
\sum_{\Delta_0}\xi_i^{(a)}D^*_iM_i\left[S''_{i,00}-\frac{M_i(S'_{i,0})^2}{e^{M_iS_i}-1}\right],
\end{flalign*}
\begin{flalign*}
\frac{\partial^2 Q\left(\bm\theta^*, \bm \xi^{(a)}\right)}{\partial \gamma_0 \partial \gamma_1}=\sum_{\Delta_1}M_iS''_{i,01}+\sum_{\Delta_1}\left[-\frac{1}{\gamma_1}+\frac{S_iS''_{i,01}-S'_{i,0}S'_{i,1}}{S_i^2}\right]+\sum_{\Delta_0}\xi_i^{(a)}D^*_iM_i\left[S''_{i,01}-\frac{M_iS'_{i,0}S'_{i,1}}{e^{M_iS_i}-1}\right],
\end{flalign*}
\begin{flalign*}
\frac{\partial^2 Q\left(\bm\theta^*, \bm \xi^{(a)}\right)}{\partial \gamma_0 \partial \gamma_{2l}}=\sum_{\Delta_1}M_iS''_{i,0(2l)}+\sum_{\Delta_1}\left[\frac{S_iS''_{i,0(2l)}-S'_{i,0}S'_{i,2l}}{S_i^2}\right]+
\sum_{\Delta_0}\xi_i^{(a)}D^*_iM_i\left[S''_{i,0(2l)}-\frac{M_iS'_{i,0}S'_{i,2l}}{e^{M_iS_i}-1}\right],
\end{flalign*}
\begin{flalign*}
\frac{\partial^2 Q\left(\bm\theta^*, \bm \xi^{(a)}\right)}{\partial \gamma_0 \partial \gamma_{3m}}=\sum_{\Delta_1}M_iS''_{i,0(3m)}+\sum_{\Delta_1}\left[\frac{S_iS''_{i,0(3m)}-S'_{i,0}S'_{i,3m}}{S_i^2}\right]+
\sum_{\Delta_0}\xi_i^{(a)}D^*_iM_i\left[S''_{i,0(3m)}-\frac{M_iS'_{i,0}S'_{i,3m}}{e^{M_iS_i}-1}\right],
\end{flalign*}
\begin{flalign*}
\frac{\partial^2 Q\left(\bm\theta^*, \bm \xi^{(a)}\right)}{\partial \gamma^2_1}=\sum_{\Delta_1}M_iS''_{i,11}+\sum_{\Delta_1}\left[\frac{\gamma_0}{\gamma^2_1}+\frac{S_iS''_{i,11}-(S'_{i,1})^2}{S_i^2}\right]+
\sum_{\Delta_0}\xi_i^{(a)}D^*_iM_i\left[S''_{i,11}-\frac{M_i(S'_{i,1})^2}{e^{M_iS_i}-1}\right],
\end{flalign*}
\begin{flalign*}
\frac{\partial^2 Q\left(\bm\theta^*, \bm \xi^{(a)}\right)}{\partial \gamma_1 \partial \gamma_{2l}}=\sum_{\Delta_1}M_iS''_{i,1(2l)}+\sum_{\Delta_1}\left[\frac{S_iS''_{i,1(2l)}-S'_{i,1}S'_{i,2l}}{S_i^2}\right]+
\sum_{\Delta_0}\xi_i^{(a)}D^*_iM_i\left[S''_{i,1(2l)}-\frac{M_iS'_{i,1}S'_{i,2l}}{e^{M_iS_i}-1}\right],
\end{flalign*}
\begin{flalign*}
\frac{\partial^2 Q\left(\bm\theta^*, \bm \xi^{(a)}\right)}{\partial \gamma_1 \partial \gamma_{3m}}=\sum_{\Delta_1}M_iS''_{i,1(3m)}+\sum_{\Delta_1}\left[\frac{S_iS''_{i,1(3m)}-S'_{i,1}S'_{i,3m}}{S_i^2}\right]+
\sum_{\Delta_0}\xi_i^{(a)}D^*_iM_i\left[S''_{i,1(3m)}-\frac{M_iS'_{i,1}S'_{i,3m}}{e^{M_iS_i}-1}\right],
\end{flalign*}
\begin{flalign*}
\frac{\partial^2 Q\left(\bm\theta^*, \bm \xi^{(a)}\right)}{\partial \gamma_{2l} \partial \gamma_{2l'}}=\sum_{\Delta_1}M_iS''_{i,(2l)(2l')}+\sum_{\Delta_1}\left[\frac{S_iS''_{i,(2l)(2l')}-S'_{i,2l}S'_{i,2l'}}{S_i^2}\right]+
\sum_{\Delta_0}\xi_i^{(a)}D^*_iM_i\left[S''_{i,(2l)(2l')}-\frac{M_iS'_{i,2l}S'_{i,2l'}}{e^{M_iS_i}-1}\right],
\end{flalign*}
\begin{flalign*}
\frac{\partial^2 Q\left(\bm\theta^*, \bm \xi^{(a)}\right)}{\partial \gamma_{2l} \partial \gamma_{3m}}=\sum_{\Delta_1}M_iS''_{i,(2l)(3m)}+\sum_{\Delta_1}\left[\frac{S_iS''_{i,(2l)(3m)}-S'_{i,2l}S'_{i,3m}}{S_i^2}\right]+
\sum_{\Delta_0}\xi_i^{(a)}D^*_iM_i\left[S''_{i,(2l)(3m)}-\frac{M_iS'_{i,2l}S'_{i,3m}}{e^{M_iS_i}-1}\right],
\end{flalign*}
\begin{flalign*}
\frac{\partial^2 Q\left(\bm\theta^*, \bm \xi^{(a)}\right)}{\partial \gamma_{3m} \partial \gamma_{3m'}}=\sum_{\Delta_1}M_iS''_{i,(3m)(3m')}+\sum_{\Delta_1}\left[\frac{S_iS''_{i,(3m)(3m')}-S'_{i,3m}S'_{i,3m'}}{S_i^2}\right]\\+
\sum_{\Delta_0}\xi_i^{(a)}D^*_iM_i\left[S''_{i,(3m)(3m')}-\frac{M_iS'_{i,3m}S'_{i,3m'}}{e^{M_iS_i}-1}\right],
\end{flalign*}
for 
$i=1, \dots, n$; $j, j'= 1, \dots, q_1$; $k, k'= 0, 1, \dots, q_2$; $r, r'= 0,1, 20, 21, \dots, 2q_2, 31, 32, \dots, 3q_1$; $l, l'= 0, 1, \dots, q_2$; $m, m'=  1, \dots, q_1$ and $x_{i0}\equiv 1.$

 
\subsection{Destructive length-biased Poisson cure model}
\hspace{5mm} We define the following quantities: $$A'_{i,j}=\frac{\partial A_i}{\partial \alpha_j}= z_{ij}\eta_i p_i F_i, \quad A'_{i,k}=\frac{\partial A_i}{\partial \beta_k}= x_{ik}\eta_i p_i (1-p_i) F_i, \quad  A'_{i,r}=\frac{\partial A_i}{\partial \gamma_r}= \eta_i p_i  F'_{i,r}, $$
$$A''_{i,jj'}=\frac{\partial^2 A_i}{\partial \alpha_j \partial \alpha_{j'}}= z_{ij} z_{ij'}\eta_i p_i F_i, \quad  A''_{i,jk}=\frac{\partial^2 A_i}{\partial \alpha_j \partial \beta_j}= x_{ik} z_{ij}\eta_i p_i (1-p_i)F_i,$$
$$A''_{i,kk'}=\frac{\partial^2 A_i}{\partial \beta_k \partial \beta_{k'}}= x_{ik} x_{ik'}\eta_i p_i (1-p_i)(1-2p_i)F_i, \quad  A''_{i,jr}=\frac{\partial^2 A_i}{\partial \alpha_j \partial \gamma_r}= z_{ij} \eta_i p_i F'_{i,r},$$
$$A''_{i,kr}=\frac{\partial^2 A_i}{\partial \beta_j \partial \gamma_r}= x_{ik} \eta_i p_i (1-p_i) F'_{i,r}, \quad A''_{i,rr'}=\frac{\partial^2 A_i}{\partial \gamma_r \partial \gamma_{r'}}=  \eta_i p_i F''_{i,rr'};$$
$$B'_{i,j}=\frac{\partial B_i}{\partial \alpha_j}= \frac{z_{ij}p_i F_i}{\eta_i e^{B_i}},\quad  B'_{i,k}=\frac{\partial B_i}{\partial \beta_k}= - \frac{x_{ik} p_i (1-p_i)}{e^{B_i}} \left[F_i+\frac{f_i}{\eta_i}\right], \quad  B'_{i,r}=\frac{\partial B_i}{\partial \gamma_r}= - e^{-B_i}\left[p_iF'_{i,r}+\frac{p_i}{\eta_i}f'_{i,r}\right],$$ 
$$B''_{i,jj'}=\frac{\partial^2 B_i}{\partial \alpha_j \partial \alpha_{j'}}= -z_{ij}z_{ij'} \frac{p_if_i(1-p_iF_i)}{\eta_i e^{2B_i}}, \quad B''_{i,kk'}=\frac{\partial^2 B_i}{\partial \beta_k \partial \beta_{k'}}= \frac{-x_{ik}x_{ik'} p_i(1-p_i)\left(1-p_i-e^{B_i}\right)\left[F_i+\frac{f_i}{\eta_i}\right]}{e^{2B_i}},$$
$$B''_{i,jk}=\frac{\partial^2 B_i}{\partial \alpha_j \partial \beta_j}= x_{ik}z_{ij}\frac{p_i(1-p_i)f_i}{ \eta_i e^{2B_i}}, \quad B''_{i,jr}=\frac{\partial^2 B_i}{\partial \alpha_j \partial \gamma_r}= 
\frac{p_iz_{ij}f'_{i,r}+p_iz_{ij}f_i\left[p_iF'_{i,r}+\frac{p_if'_{i,r}}{\eta_i}\right]}{\eta_ie^{B_i}},$$
$$B''_{i,kr}=\frac{\partial^2 B_i}{\partial \beta_j \partial \gamma_r}= -\frac{x_{ik}p_i(1-p_i)\left[F'_{i,r}+\frac{f'_{i,r}}{\eta_i}\right]}{e^{B_i}}-\frac{x_{ik}p_i(1-p_i)\left[F_i+\frac{f_i}{\eta_i}\right]\left[p_iF'_{i,r}+\frac{p_if'_{i,r}}{\eta_i}\right]}{e^{2B_i}},$$
$$B''_{i,rr'}=\frac{\partial^2 B_i}{\partial \gamma_r \partial \gamma_{r'}}=  -\frac{p_i\left[F''_{i,rr'}+\frac{f''_{i,rr'}}{\eta_i}\right]}{e^{B_i}}-\frac{p^2_i\left[F'_{i,r}+\frac{f'_{i,r}}{\eta_i}\right]\left[F'_{i,r'}+\frac{f'_{i,r'}}{\eta_i}\right]}{e^{2B_i}};$$
$$C'_{i,j}=\frac{\partial C_i}{\partial \alpha_j}= \frac{z_{ij}\eta_ip_i(1-F_i)}{e^{-\eta_ip_i(1-F_i)}}, \quad C'_{i,k}=\frac{\partial C_i}{\partial \beta_k}= \frac{x_{ik}\eta_ip_i(1-p_i)(1-F_i)}{e^{-\eta_ip_i(1-F_i)}}, \quad C'_{i,r}=\frac{\partial C_i}{\partial \gamma_r}= - \frac{\eta_ip_iF'_{i,r}}{e^{-\eta_ip_i(1-F_i)}}, $$
$$C''_{i,jj'}=\frac{\partial^2 C_i}{\partial \alpha_j \partial \alpha_{j'}}= \frac{z_{ij}z_{ij'}\eta_ip_i(1-F_i)}{e^{-\eta_ip_i(1-F_i)}[1+\eta_ip_i(1-F_i)]^{-1}}, \quad C''_{i,jk}=\frac{\partial^2 C_i}{\partial \alpha_j \partial \beta_j}=\frac{x_{ik}z_{ij}\eta_ip_i(1-p_i)(1-F_i)}{e^{-\eta_ip_i(1-F_i)}[1+\eta_ip_i(1-F_i)]^{-1}},$$
$$C''_{i,kk'}=\frac{\partial^2 C_i}{\partial \beta_k \partial \beta_{k'}}= \frac{x_{ik}x_{ik'}\eta_ip_i(1-p_i)(1-F_i){e^{\eta_ip_i(1-F_i)}}}{[1-2p_i+\eta_ip_i(1-p_i)(1-F_i)]^{-1}}, \quad C''_{i,jr}=\frac{\partial^2 C_i}{\partial \alpha_j \partial \gamma_r}= \frac{-z_{ij}\eta_ip_iF'_{i,r}}{e^{-\eta_ip_i(1-F_i)}[1+\eta_ip_i(1-F_i)]^{-1}},$$
$$C''_{i,kr}=\frac{\partial^2 C_i}{\partial \beta_j \partial \gamma_r}= -\frac{x_{ik}\eta_ip_i(1-p_i)F'_{i,r}}{e^{-\eta_ip_i(1-F_i)}[1+\eta_ip_i(1-F_i)]^{-1}}, \quad C''_{i,rr'}=\frac{\partial^2 C_i}{\partial \gamma_r \partial \gamma_{r'}}=  -\frac{\eta_ip_i\left(F''_{i,rr'}-\eta_ip_iF'_{i,r}F'_{i,r'}\right)}{e^{-\eta_ip_i(1-F_i)}};$$
$$D'_{i,j}=\frac{\partial D_i}{\partial \alpha_j}= 0, \quad D'_{i,k}=\frac{\partial D_i}{\partial \beta_k}= \frac{x_{ik}p_i(1-F_i)}{1-p_i}, \quad D'_{i,r}=\frac{\partial D_i}{\partial \gamma_r}= -\frac{p_iF'_{i,r}}{1-p_i},$$ 
$$D''_{i,jj'}=\frac{\partial^2 D_i}{\partial \alpha_j \partial \alpha_{j'}}=0, \quad D''_{i,jk}=\frac{\partial^2 D_i}{\partial \alpha_j \partial \beta_j}=0, \quad D''_{i,kk'}=\frac{\partial^2 D_i}{\partial \beta_k \partial \beta_{k'}}= 0,$$
$$D''_{i,jr}=\frac{\partial^2 D_i}{\partial \alpha_j \partial \gamma_r}= 
\frac{x_{ik}x_{ik'}p_i(1-F_i)}{1-p_i}, \quad D''_{i,kr}=\frac{\partial^2 D_i}{\partial \beta_j \partial \gamma_r}= -\frac{x_{ik}p_iF'_{i,r}}{1-p_i}, \quad D''_{i,rr'}=\frac{\partial^2 D_i}{\partial \gamma_r \partial \gamma_{r'}}=  -\frac{p_iF''_{i,rr'}}{1-p_i},$$
for 
$i=1, \dots, n$; $j, j'= 1, \dots, q_1$; $k, k'= 0, 1, \dots, q_2$; $r, r'= 0,1, 20, 21, \dots, 2q_2, 31, 32, \dots, 3q_1$; $l, l'= 0, 1, \dots, q_2$; $m, m'=  1, \dots, q_1$ and $x_{i0}\equiv 1.$\\

\hspace{5mm}Then, we have: 
\begin{flalign*}
 \frac{\partial Q\left(\bm\theta^*, \bm \xi^{(a)}\right)}{\partial \alpha_j}=\sum_{\Delta_1} z_{ij}-\sum_{\Delta_1}A'_{i,j}
+\sum_{\Delta_1}B'_{i,j}-\sum_{\Delta_0}z_{ij}\eta_i p_i +\sum_{\Delta_0} \xi_i^{(a)} \frac{C'_{i,j}D_i}{C_i D_i-1},
\end{flalign*}
\begin{flalign*}
 \frac{\partial Q\left(\bm\theta^*, \bm \xi^{(a)}\right)}{\partial \beta_k}=\sum_{\Delta_1}x_{ik}(1-p_i)-\sum_{\Delta_1}A'_{i,k}
+\sum_{\Delta_1}B'_{i,k}-\sum_{\Delta_0}x_{ik}\eta_i p_i(1-p_i) + \sum_{\Delta_0}x_{ik}p_i+\sum_{\Delta_0} \xi_i^{(a)} \frac{C'_{i,k}D_i+D'_{i,k}C_i}{C_iD_i-1},
\end{flalign*}
\begin{flalign*}
 \frac{\partial Q\left(\bm\theta^*, \bm \xi^{(a)}\right)}{\partial \gamma_r}=\sum_{\Delta_1}\frac{\partial \log f_i}{\partial \gamma_r}-\sum_{\Delta_1}A'_{i,r}+\sum_{\Delta_1}B'_{i,r} +\sum_{\Delta_0} \xi_i^{(a)} \frac{C'_{i,r}D_i+D'_{i,r}C_i}{C_iD_i-1},
\end{flalign*}
\begin{flalign*}
 \frac{\partial^2 Q\left(\bm\theta^*, \bm \xi^{(a)}\right)}{\partial \alpha_j \partial \alpha_j'}=-\sum_{\Delta_1}A''_{i,jj'}
+\sum_{\Delta_1}B''_{i,jj'}-\sum_{\Delta_0}z_{ij}z_{ij'}\eta_i p_i +\sum_{\Delta_0} \xi_i^{(a)} D_i\left[\frac{D_i(C_i C''_{i,jj'}-C'_j C'_{j'})-C''_{i,jj'}}{(C_iD_i-1)^2}\right],
\end{flalign*}
\begin{flalign*}
 \frac{\partial^2 Q\left(\bm\theta^*, \bm \xi^{(a)}\right)}{\partial \alpha_j \partial \beta_k}=-\sum_{\Delta_1}A''_{i,jk}
+\sum_{\Delta_1}B''_{i,jk}+\sum_{\Delta_0} \xi_i^{(a)} D_i\left[\frac{D_i(C_i C''_{i,jk}-C'_j C'_{k})-C''_{i,jk}}{(C_iD_i-1)^2}\right],
\end{flalign*}
\begin{flalign*}
 \frac{\partial^2 Q\left(\bm\theta^*, \bm \xi^{(a)}\right)}{\partial \alpha_j \partial \gamma_r}=-\sum_{\Delta_1}A''_{i,jr}
+\sum_{\Delta_1}B''_{i,jr}+\sum_{\Delta_0} \xi_i^{(a)} D_i\left[\frac{D_i(C_i C''_{i,jr}-C'_j C'_{r})-C''_{i,jr}}{(C_iD_i-1)^2}\right],
\end{flalign*}
\begin{align*}
 \frac{\partial^2 Q\left(\bm\theta^*, \bm \xi^{(a)}\right)}{\partial \beta_k \partial \beta_{k'}}&=-\sum_{\Delta_1} x_{ik} x_{ik'}p_i(1-p_i)
-\sum_{\Delta_1}A''_{i,kk'}+\sum_{\Delta_1}B''_{i,kk'}+\sum_{\Delta_0} x_{ik} x_{ik'}p_i(1-p_i)\\&
 -\sum_{\Delta_1}x_{ik} x_{ik'}\eta_ip_i(1-p_i)(1-2p_i)-\sum_{\Delta_0} \xi_i^{(a)}\left[\frac{\{C_iD''_{i,kk'}+D_iC''_{i,kk'}+C'_{i,k}D'_{i,k'}+C'_{i,k'}D'_{i,k}\}}{(C_iD_i-1)^2}\right]\\&
+\sum_{\Delta_0} \xi_i^{(a)} \left[\frac{D^2_i\{C_iC''_{i,kk'}-C'_{i,k}C'_{i,k'}\}+C^2_i\{D_iD''_{i,kk'}-D'_{i,k}D'_{i,k'}\}}{(C_iD_i-1)^2}\right],
\end{align*}
\begin{align*}
 \frac{\partial^2 Q\left(\bm\theta^*, \bm \xi^{(a)}\right)}{\partial \beta_k \partial \gamma_{k}}&=
-\sum_{\Delta_1}(A''_{i,kr}-B''_{i,kr})-\sum_{\Delta_0} \xi_i^{(a)}\left[\frac{\{C_iD''_{i,kr}+D_iC''_{i,kr}+C'_{i,k}D'_{i,r}+C'_{i,r}D'_{i,k}\}}{(C_iD_i-1)^2}\right]
\\&+\sum_{\Delta_0} \xi_i^{(a)}
 \left[\frac{D^2_i\{C_iC''_{i,kr}-C'_{i,k}C'_{i,r}\}+C^2_i\{D_iD''_{i,kr}-D'_{i,k}D'_{i,r}\}}{(C_iD_i-1)^2}\right],
\end{align*}
and
\begin{align*}
 \frac{\partial^2 Q\left(\bm\theta^*, \bm \xi^{(a)}\right)}{\partial \gamma_r \partial \gamma_{r'}}&=\sum_{\Delta_1} \frac{\partial^2 \log f_i}{\partial \gamma_r \partial \gamma_{r'}}
-\sum_{\Delta_1}A''_{i,rr'}
+\sum_{\Delta_1}B''_{i,rr'}-\sum_{\Delta_0} \xi_i^{(a)}\left[\frac{\{C_iD''_{i,rr'}+D_iC''_{i,rr'}+C'_{i,r}D'_{i,r'}+C'_{i,r'}D'_{i,r}\}}{(C_iD_i-1)^2}\right] \\&
+\sum_{\Delta_0} \xi_i^{(a)}
 \left[\frac{D^2_i\{C_iC''_{i,rr'}-C'_{i,r}C'_{i,r'}\}+C^2_i\{D_iD''_{i,rr'}-D'_{i,r}D'_{i,r'}\}}{(C_iD_i-1)^2}\right].
\end{align*}

\subsection{Destructive negative binomial cure model}
\hspace{5mm} Let us define the following quantities: $$G'_{i,j}=\frac{\partial G_i}{\partial \alpha_j}=\frac{z_{ij}E_i(F_i-1)}{(1+E_i)^2}, \quad G'_{i,k}=\frac{\partial G_i}{\partial \beta_k}=\frac{x_{ik}E_i(1-p_i)(F_i-1)}{(1+E_i)^2}, \quad G'_{i,r}=\frac{\partial G_i}{\partial \gamma_r}=\frac{E_i F'_{i,r}}{(1+E_i)},$$
$$ G''_{i,jj'}=\frac{\partial^2 G_i}{\partial \alpha_j \partial \alpha_{j'}}=\frac{z_{ij}z_{ij'}E_i(1-E_i)(F_i-1)}{(1+E_i)^3}, \quad  G''_{i,jk}=\frac{\partial^2 G_i}{\partial \alpha_j \partial \beta_k}=\frac{z_{ij}x_{ik}E_i(1-p_i)(1-E_i)(F_i-1)}{(1+E_i)^3},$$
$$ G''_{i,kk'}=\frac{\partial^2 G_i}{\partial \beta_k \partial \beta_{k'}}=\frac{x_{ik}x_{ik'}E_i(1-p_i)^2(1-E_i)(F_i-1)}{(1+E_i)^3}, \quad G''_{i,jr}=\frac{\partial^2 G_i}{\partial \alpha_j \partial \gamma_r}=\frac{z_{ij}E_iF'_{i,r}}{(1+E_i)^2},$$ 
$$ G''_{i,kr}=\frac{\partial^2 G_i}{\partial \beta_k \partial \gamma_r}=\frac{x_{ik}E_i(1-p_i)F'_{i,r}}{(1+E_i)^2},\text{ and } G''_{i,rr'}=\frac{\partial^2 G_i}{\partial \gamma_r \partial \gamma_{r'}}=\frac{E_i}{(1+E_i)}F''_{i,rr'},$$
where 
$i=1, \dots, n$; $j, j'= 1, \dots, q_1$; $k, k'= 0, 1, \dots, q_2$; $r, r'= 0,1, 20, 21, \dots, 2q_2, 31, 32, \dots, 3q_1$; $l, l'= 0, 1, \dots, q_2$; $m, m'=  1, \dots, q_1$ and $x_{i0}\equiv 1.$\\

\hspace{5mm} Then, we have:
\begin{flalign*}
\frac{\partial Q\left(\bm\theta^*, \bm \xi^{(a)}\right)}{\partial \alpha_j}=\sum_{\Delta_1}z_{ij}-\left(\frac{1}{\phi}+1\right)\sum_{\Delta_1}z_{ij}\frac{E_iF_i}{1+E_iF_i} - \frac{1}{\phi}\sum_{\Delta_0}z_{ij}\frac{E_i}{1+E_i}+\sum_{\Delta_0}\xi_i^{(a)}\frac{G'_{i,j}}{\phi G_i ( G_i^{1/\phi}-1)},
\end{flalign*}
\begin{flalign*}
\frac{\partial Q\left(\bm\theta^*, \bm \xi^{(a)}\right)}{\partial \beta_k}=\sum_{\Delta_1}x_{ik}(1-p_i)-\left(\frac{1}{\phi}+1\right)\sum_{\Delta_1}x_{ik}\frac{E_iF_i (1-p_i)}{1+E_iF_i} - \frac{1}{\phi}\sum_{\Delta_0}x_{ik}\frac{E_i (1-p_i)}{1+E_i}+\sum_{\Delta_0}\xi_i^{(a)}\frac{G'_{i,k}}{\phi G_i ( G_i^{1/\phi}-1)},
\end{flalign*}
\begin{flalign*}
\frac{\partial Q\left(\bm\theta^*, \bm \xi^{(a)}\right)}{\partial \gamma_r}=-\left(\frac{1}{\phi}+1\right)\sum_{\Delta_1}\frac{E_i F'_{i,r}}{1+E_iF_i} + \sum_{\Delta_1}\frac{\partial \log f_i}{\partial \gamma_r}  +\sum_{\Delta_0}\xi_i^{(a)}\frac{G'_{i,r}}{\phi G_i ( G_i^{1/\phi}-1)},
\end{flalign*}
\begin{flalign*}
 \frac{\partial^2 Q\left(\bm\theta^*, \bm \xi^{(a)}\right)}{\partial \alpha_j \partial \alpha_{j'}}&=-\left(\frac{1}{\phi}+1\right)\sum_{\Delta_1}\frac{z_{ij}z_{ij'}E_iF_i}{(1+E_iF_i)^2}-\frac{1}{\phi}\sum_{\Delta_0}\frac{z_{ij}z_{ij'}E_i}{(1+E_i)^2}\\&+\sum_{\Delta_0} \xi_i^{(a)}\left[\frac{G''_{i,jj'} G_i ( G_i^{1/\phi}-1)-G'_{i,j}G'_{i,j'}\{(1/\phi+1)G_i^{1/\phi}-1\}}{\phi \left(G_i^{1/\phi+1}-1\right)^2}\right],
\end{flalign*}
\begin{flalign*}
 \frac{\partial^2 Q\left(\bm\theta^*, \bm \xi^{(a)}\right)}{\partial \alpha_j \partial \beta_k}=-\left(\frac{1}{\phi}+1\right)\sum_{\Delta_1}\frac{z_{ij}x_{ik}E_iF_i(1-p_i)}{(1+E_iF_i)^2}-\frac{1}{\phi}\sum_{\Delta_0}\frac{z_{ij}x_{ik}(1-p_i)E_i}{(1+E_i)^2}&\\+\sum_{\Delta_0} \xi_i^{(a)}\left[\frac{G''_{i,jk} G_i ( G_i^{1/\phi}-1)-G'_{i,j}G'_{i,k}\{(1/\phi+1)G_i^{1/\phi}-1\}}{\phi \left(G_i^{1/\phi+1}-1\right)^2}\right],
\end{flalign*}
\begin{flalign*}
 \frac{\partial^2 Q\left(\bm\theta^*, \bm \xi^{(a)}\right)}{\partial \beta_k \partial \beta_{k'}}=&-\left(\frac{1}{\phi}+1\right)\sum_{\Delta_1}\frac{x_{ik}x_{ik'}E_iF_i(1-p_i)(1-2p_i-E_iF_ip_i)}{(1+E_iF_i)^2}-\frac{1}{\phi}\sum_{\Delta_0}\frac{x_{ik}x_{ik'}(1-p_i)^2E_i}{(1+E_i)^2}\\&+\sum_{\Delta_0} \xi_i^{(a)}\left[\frac{G''_{i,kk'} G_i ( G_i^{1/\phi}-1)-G'_{i,k}G'_{i,k'}\{(1/\phi+1)G_i^{1/\phi}-1\}}{\phi \left(G_i^{1/\phi+1}-1\right)^2}\right],
\end{flalign*}
\begin{flalign*}
 \frac{\partial^2 Q\left(\bm\theta^*, \bm \xi^{(a)}\right)}{\partial \alpha_j \partial \gamma_r}=-\left(\frac{1}{\phi}+1\right)\sum_{\Delta_1}\frac{z_{ij}E_iF'_{i,r}}{(1+E_iF_i)^2}+\sum_{\Delta_0} \xi_i^{(a)}\left[\frac{G''_{i,jr} G_i ( G_i^{1/\phi}-1)-G'_{i,j}G'_{i,r}\{(1/\phi+1)G_i^{1/\phi}-1\}}{\phi \left(G_i^{1/\phi+1}-1\right)^2}\right],
\end{flalign*}
\begin{flalign*}
 \frac{\partial^2 Q\left(\bm\theta^*, \bm \xi^{(a)}\right)}{\partial \beta_k \partial \gamma_r}=-\left(\frac{1}{\phi}+1\right)\sum_{\Delta_1}\frac{x_{ik}(1-p_i)E_iF'_{i,r}}{(1+E_iF_i)^2}+\sum_{\Delta_0} \xi_i^{(a)}\left[\frac{G''_{i,kr} G_i ( G_i^{1/\phi}-1)-G'_{i,k}G'_{i,r}\{(1/\phi+1)G_i^{1/\phi}-1\}}{\phi \left(G_i^{1/\phi+1}-1\right)^2}\right],
\end{flalign*}
and
\begin{flalign*}
\frac{\partial^2 Q\left(\bm\theta^*, \bm \xi^{(a)}\right)}{\partial \gamma_r \partial \gamma_{r'}}&=-\left(\frac{1}{\phi}+1\right)\sum_{\Delta_1}\left[\frac{E_iF''_{i,rr'}+E_i^2\{F_iF''_{i,rr'}-F'_{i,r}F'_{i,r'}\}}{(1+E_iF_i)^2}\right]+\sum_{\Delta_1}\frac{\partial^2 \log f_i}{\partial \gamma_r \partial \gamma_{r'}}\\&+\sum_{\Delta_0} \xi_i^{(a)}\left[\frac{G''_{i,rr'} G_i ( G_i^{1/\phi}-1)-G'_{i,r}G'_{i,r'}\{(1/\phi+1)G_i^{1/\phi}-1\}}{\phi \left(G_i^{1/\phi+1}-1\right)^2}\right].
\end{flalign*}

\end{document}